\documentclass[prd,aps,showpacs,nofootinbib]{revtex4}%
\usepackage{graphicx}
\usepackage{amsmath}
\usepackage{amsfonts}
\usepackage{amssymb}%
\newcommand{\be}{\begin{equation}}
\newcommand{\ee}{\end{equation}}
\newcommand{\bq}{\begin{eqnarray}}
\newcommand{\eq}{\end{eqnarray}}
\begin{document}
\title{\textbf{Noncommutative Conformally Coupled Scalar Field Cosmology and its
Commutative Counterpart}}
\author{G. D. Barbosa}
\email{gbarbosa@cbpf.br}
\affiliation{Centro Brasileiro de Pesquisas F\'{\i}sicas, CBPF, Rua Dr. Xavier Sigaud 150,
22290-180, Rio de Janeiro, Brazil }

\begin{abstract}
We study the implications of a\ noncommutative geometry of the minisuperspace
variables for the FRW universe with a conformally coupled scalar field. The
investigation is carried out by means of a comparative study of the universe
evolution in four different scenarios: classical commutative, classical
noncommutative, quantum commutative, and quantum noncommutative, the last two
employing the Bohmian formalism of quantum trajectories. The role of
noncommutativity is discussed by drawing a parallel between its realizations
in two possible frameworks for physical interpretation: the NC-frame, where it
is manifest in the universe degrees of freedom, and in the C-frame, where it
is manifest through $\theta$-dependent terms in the Hamiltonian. As a result
of our comparative analysis, we find that noncommutative geometry can remove
singularities in the classical context for sufficiently large values of
$\theta$. Moreover, under special conditions, the classical noncommutative
model can admit bouncing solutions characteristic of the commutative quantum
FRW universe. In the quantum context, we find non-singular universe solutions
containing bounces or being periodic in the quantum commutative model. When
noncommutativity effects are turned on in the quantum scenario, they can
introduce significant modifications that change the singular behavior of the
universe solutions or\ that render them dynamical whenever they are static in
the commutative case. The effects of noncommutativity are completely specified
only when one of the frames for its realization is adopted as the physical
one. Non-singular solutions in the NC-frame can be mapped into singular ones
in the C-frame.

\end{abstract}
\pacs{98.80.Qc,04.60.Kz,11.10.Nx,11.10.Lm}
\maketitle

\section{Introduction}

Over the latest years a great deal of work and effort has been done in the
direction of understanding canonical noncommutative field theories and quantum
mechanics (see \cite{1,2} and references therein). The recent interest in
these theories is motivated by works that establish a connection between
noncommutative geometry and string theory \cite{3}. Intensive research is
carried out to investigate their interesting properties, such as the IR-UV
mixing and nonlocality \cite{5}, Lorentz violation \cite{6}, new physics at
very short distances \cite{1,2}, and the equivalence between translations in
the noncommutative directions and gauge transformations (see,
e.g.\ \cite{2,6.5}).

Several investigations have been pursued to verify the possible role of
noncommutativity in a great deal of cosmological scenarios. Among them we
quote Newtonian cosmology \cite{11}, cosmological perturbation theory and
inflationary cosmology \cite{12}, noncommutative gravity \cite{14}, and
quantum cosmology \cite{15,16}. In a previous work \cite{16}, an investigation
into the influence of noncommutativity of the minisuperspace variables in the
early universe scenario was carried out for the Kantowski-Sachs universe.
Although noncommutativity effects proved to be relevant to the universe
history at intermediate times, they were shown not to be capable of removing
the future and past cosmological singularities of that model in the classical
context. In the quantum context, on the other hand, non-singular universe
solutions were shown to be present. However, since they exist for the
commutative quantum Kantowski-Sachs universe, their presence in the ensemble
of solutions of the noncommutative quantum model cannot be attributed to the
noncommutativity effects.

Although the investigation carried out in \cite{16} was restricted to a
particular model, one expected that some of the results obtained there could
be of general validity. Indeed, in this work we shall show that
noncommutativity can appreciably modify the evolution of the
Friedman-Robertson-Walker (FRW) universe with a conformally coupled scalar
field \cite{17,18}.\ As in reference \cite{16}, our investigation is carried
out by means of a comparative study of the universe evolution in four
different scenarios: classical commutative, classical noncommutative, quantum
commutative and quantum noncommutative.\ The main motivation for the choice of
the conformally coupled scalar field is that it admits exact solutions in the
simpler cases discussed along this work and it is rich enough to be useful as
a probe for the significant modifications noncommutative geometry introduces
in classical and quantum cosmologies. The analytical treatment renders easy
the study of the singular behavior of the model in its four versions. As we
shall show later, even in the classical context noncommutative geometry can
remove singularities. Moreover, depending on the value of the noncommutative
parameter, noncommutative classical models can mimic quantum effects.

As an interpretation for quantum theory, we are adopting the Bohmian one,
which we briefly review in section 4.\ Originally proposed by Bohm\ in 1952
\cite{19}, and further developed by him in a collaboration with Hiley
\cite{19.2}, such an interpretation of quantum theory has acquired an
increasing number of adepts along the years \cite{19.3}. Decisive
contributions for the development of the\ Bohmian quantum physics were made by
Bell \cite{19.4}, who was its arduous defender for three decades, by Holland
\cite{20}, and by\ D. D\"{u}rr, S. Goldstein and N. Zangh\`{\i} (see, e.g.,
\cite{20.3}-\cite{20.7} and ref. therein). Due to the capability it has to
reproduce the experimental results of quantum mechanics and quantum field
theory providing an intuitive interpretation of the underlying dynamics,
Bohmian quantum physics is presently an issue of interest for broad community
(see, e.g., \cite{21}). In this work the main reason that compels us to adopt
the Bohmian interpretation in quantum cosmology is the absence of external
observers in the primordial quantum universe, which renders the standard
Copenhagen interpretation inapplicable in its description. As an alternative,
the Bohmian interpretation has been employed\ in several works of quantum
cosmology and\ quantum gravity\ (see, e.g., \cite{16,18,22}).
Other\ interesting\ aspect of the Bohmian approach to quantum theory, which
motivates us for its adoption,\ is the efficient framework it provides for
comparison between the classical and quantum counterparts of a physical\ model
in the common language of trajectories. We shall benefit from this facility in
our comparative study of the four versions of the FRW universe. Since our
preference for Bohmian interpretation is justified for technical, rather than
philosophical reasons, we shall not focus our discussion on fundamental
questions regarding the interpretation of quantum theory (for references on
this subject see \cite{18,22.1}). Instead, whenever possible,\ we shall give
preference for the interpretation-independent aspects.

This work is organized as follows. Sections 2 and 3\ are devoted to a
comparative study of the classical\ FRW with a conformally coupled scalar
field and its noncommutative counterpart. In section 4 we present the
commutative\ quantum version of the model, and analyze it by using the Bohmian
formalism of quantum trajectories. A similar study is carried out in sections
5 and 6, which are concerned with the noncommutative quantum version of the
model. In section 7 we end up with a general discussion and summary of the
main results.

\section{The Conformally Coupled Scalar Field Model}

As a reference for the identification of the noncommutative effects later, it
is interesting to consider first the commutative classical FRW universe, which
we describe as follows. We shall restrict our considerations to the case of
constant positive curvature of the spatial sections. The action for the
conformally coupled scalar field model in this case\ is \cite{18}
\begin{equation}
S=\int d^{4}x\sqrt{-g}\left[  -\frac{1}{2}\varphi_{;\mu}\varphi^{;\mu}%
+\frac{1}{16\pi G}R-\frac{1}{12}R\varphi^{2}\right]  , \label{9}%
\end{equation}
where $g_{\mu\nu}$ is the four-metric, $g$ its determinant, $R$ is the scalar
curvature, and $\varphi$ is the scalar field. Units are chosen such that
$\hbar=c=1$ and $8\pi G=3l_{p}^{2}$, where $l_{p}$ is the Planck length. For
the FRW model with an homogeneous scalar field the following \textit{ansatz}
of minisuperspace can be adopted%

\begin{equation}
\left\{
\begin{array}
[c]{c}%
ds^{2}=-N^{2}(t)dt^{2}+a^{2}(t)\left[  \frac{dr^{2}}{1-r^{2}}+r^{2}\left(
d\theta^{2}+\sin^{2}\theta d\varphi^{2}\right)  \right]  ,\\
\\
\varphi=\varphi(t).\hspace{7.2cm}%
\end{array}
\right.  \label{10}%
\end{equation}

By substituting (\ref{10}) in (\ref{9}) and rescaling the scalar field as
$\chi=\varphi al_{p}/\sqrt{2,}$ we have the following
minisuperspace\ action\footnote{We have discarded total time derivatives and
\ integrated out the spatial degrees of freedom since they are not relevant
for the equations of motion.}
\begin{equation}
S=\int dt\left(  Na-\frac{a\dot{a}^{2}}{N}+\frac{a\dot{\chi}^{2}}{N}%
-\frac{N\chi^{2}}{a}\right)  . \label{11}%
\end{equation}
The corresponding Hamiltonian is
\begin{equation}
H=N\left[  -\frac{P_{a}^{2}}{4a}+\frac{P_{\chi}^{2}}{4a}-a+\frac{\chi^{2}}%
{a}\right]  =N\mathcal{H}, \label{12}%
\end{equation}
where the canonical momenta are%
\begin{equation}
P_{a}=-\frac{2a\dot{a}}{N}\text{, \ \ }P_{\chi}=\frac{2a\dot{\chi}}{N}.
\label{12.5}%
\end{equation}
As the Poisson Brackets for the classical phase space variables we have
\begin{equation}
\left\{  a,\chi\right\}  =0,\text{ \ }\left\{  a,P_{a}\right\}  =1\text{,
\ }\left\{  \chi,P_{\chi}\right\}  =1\text{,\ \ }\left\{  P_{a},P_{\chi
}\right\}  =0. \label{13}%
\end{equation}
The equations of motion for the metric and matter field variables $a$,
$P_{a},$ $\chi$ and $P_{\chi}$\ that follow from (\ref{12}) and (\ref{13}) are%

\begin{equation}
\left\{
\begin{array}
[c]{c}%
\dot{a}=\left\{  a,H\right\}  =-NP_{a}/2a,\\
\\
\dot{P}_{a}=\left\{  P_{a},H\right\}  =2N,\hspace{1cm}\\
\\
\dot{\chi}=\left\{  \chi,H\right\}  =NP_{\chi}/2a,\hspace{0.2cm}\\
\\
\dot{P}_{\chi}=\left\{  P_{\chi},H\right\}  =-2N\chi/a.
\end{array}
\right.  \label{14}%
\end{equation}
From now on we shall adopt conformal time gauge $N=a.$\ The general solution
of (\ref{14}) for $a$ and $\chi$ in this gauge is
\begin{equation}
a(t)=\left(  A+C\right)  \cos(t)+\left(  B+D\right)  \sin(t), \label{16}%
\end{equation}%
\begin{equation}
\chi(t)=\left(  A-C\right)  \cos(t)+\left(  B-D\right)  \sin(t), \label{17}%
\end{equation}
where the super-Hamiltonian constraint $\mathcal{H}\approx0$ imposes the
relation
\begin{equation}
AC+BD=0. \label{18}%
\end{equation}

As it can be seen, the classical commutative solutions are necessarily
singular in the past and in the future. Figs. $1(a),(b),(c)$ and $(d)$ present
plots of the solution for $a(t)$ in the dashed curves for given values of
$A,B,$ and $C$ [$D$ is fixed by (\ref{18})].

\section{Noncommutative Deformation of the Classical Model}

Let us\ introduce a noncommutative classical geometry in our universe model by
keeping the Hamiltonian with the same functional form as (\ref{12}), but now
valued on noncommutative variables,
\begin{equation}
H=N\left[  -\frac{P_{a_{nc}}^{2}}{4a_{nc}}+\frac{P_{\chi_{nc}}^{2}}{4a_{nc}%
}-a_{nc}+\frac{\chi_{nc}^{2}}{a_{nc}}\right]  , \label{19}%
\end{equation}
where $a_{nc}$, $\chi_{nc}$, $P_{a_{nc}}$ and $P_{\chi_{nc}}$ satisfy the
deformed Poisson brackets%

\begin{equation}
\left\{  a_{nc},\chi_{nc}\right\}  =\theta,\text{ \ }\left\{  a_{nc}%
,P_{a_{nc}}\right\}  =1\text{, \ }\left\{  \chi_{nc},P_{\chi_{nc}}\right\}
=1\text{,\ \ }\left\{  P_{a_{nc}},P_{\chi_{nc}}\right\}  =0. \label{20}%
\end{equation}
By making the substitution
\begin{equation}
a_{nc}=a_{c}-\frac{\theta}{2}P_{\chi_{c}}\text{, \ }\chi_{nc}=\chi
\text{$_{c}+$}\frac{\theta}{2}P_{a_{c}}\text{,\ \ }P_{a_{c}}=P_{a_{nc}%
}\text{,\ \ }P_{\chi_{c}}=P_{\chi_{nc}}, \label{21}%
\end{equation}
the theory defined by (\ref{19}) and (\ref{20})\ can be mapped into a theory
where the metric and matter variables\ satisfy the Poisson brackets
\begin{equation}
\left\{  a_{c},\chi_{c}\right\}  =0,\text{ \ }\left\{  a_{c},P_{a_{c}%
}\right\}  =1\text{,\ \ }\left\{  \chi_{c},P_{\chi_{c}}\right\}
=1\text{,\ \ }\left\{  P_{a_{c}},P_{\chi_{c}}\right\}  =0. \label{22}%
\end{equation}

Written in terms of $a_{c}$, $\chi_{c}$, $P_{a_{c}}$ and $P_{\chi_{c}},$\ the
Hamiltonian (\ref{19}) exhibits the noncommutative content of the theory
through $\theta$-dependent terms. Two distinct physical theories, one
considering $a_{c}$ and $\chi_{c}$,\ and the other considering $a_{nc}$ and
$\chi_{nc}$ as the physical scale factor and matter field can be assumed to
emerge from (\ref{19}), (\ref{20}), (\ref{21}) and (\ref{22}).\ In the case
where $a_{c}$ and $\chi_{c}$ are assumed as the preferred variables for
physical interpretation, the theory\ can be interpreted as a \textquotedblleft
commutative\textquotedblright\ one with a modified interaction. We shall refer
to this theory as being realized in the \textquotedblleft
C-frame\textquotedblright. To the other possible theory, which assumes
$a_{nc}$ and $\chi_{nc}$ as the constituents of the physical metric and matter
field, we shall refer as realized in the \textquotedblleft
NC-frame\textquotedblright.\footnote{The situation here is, in a certain
sense, similar to that of\ generalized theories of gravitation. There are two
different frames for physical interpretation, as the Einstein frame and the
Jordan frame \cite{22.5}. There is a mapping between them, as the conformal
transformation that maps the the Jordan into the Einstein frame. However, as
in the\ generalized theories of gravitation, the noncommutative cosmologies in
the different frames are not physically equivalent. The reason is the same as
there: the theories are not the same because the physical objects they refer
to are not identical.} There are works that privilege the C-frame approach
(e.g. \cite{15}), and others the NC-frame (e.g. \cite{16,22.2,22.3}). Some
works rely on the assumption that the difference\ between C- and
NC-variables\ is negligible (e.g. \cite{22.7}). However, as showed in
\cite{22.3}, even in simple models the difference in behavior between these
two types of variables can be appreciable. In what follows we will show that
in the cosmological scenario the assumption of the NC- or C-frame point of
view as preferential for physical interpretation leads to dramatic differences
in the analysis of the universe history.\ A parallel between the theories in
both frames realizations is draw in the classical context here, and in the
quantum context in section 6.

Methodologically, the route we shall follow in the computation of the
configuration variables in this classical context is the same as that adopted
in the noncommutative quantum\ case discussed later on. We shall depart from
the C-frame and calculate $a_{c}(t)$, $\chi_{c}(t)$, $P_{a_{c}}(t),$ and
$P_{\chi_{c}}(t)$. After that, we shall use (\ref{21}) to obtain the
corresponding $a_{nc}(t)$ and $\chi_{nc}(t)$ in the NC-frame. The computation
of the physical quantities is rendered simpler by the gauge choice $N=a_{nc}$,
which from now on will be assumed as the gauge employed in all calculations.

The\ equations of motion for the variables $a_{c}(t)$, $\chi_{c}(t)$,
$P_{a_{c}}(t)$ and $P_{\chi_{c}}(t)$ are
\begin{equation}
\left\{
\begin{array}
[c]{c}%
\dot{a}_{c}=\left\{  a_{c},H\right\}  =-\frac{1}{2}\left(  1-\theta
^{2}\right)  P_{a_{c}}+\theta\chi_{c},\\
\\
\dot{P}_{a_{c}}=\left\{  P_{a_{c}},H\right\}  =2a_{c}-\theta P_{\chi_{c}%
},\hspace{1.7cm}\\
\\
\dot{\chi}_{c}=\left\{  \chi_{c},H\right\}  =\frac{1}{2}\left(  1-\theta
^{2}\right)  P_{\chi_{c}}+\theta a_{c},\hspace{0.3cm}\\
\\
\dot{P}_{\chi_{c}}=\left\{  P_{\chi_{c}},H\right\}  =-2\chi_{c}-\theta
P_{a_{c}}.\hspace{1.3cm}%
\end{array}
\right.  \label{23}%
\end{equation}
According to the values of $\theta$, the system (\ref{23}) allows three types
of solutions, which we describe below.

\subsubsection{Case $\left|  \theta\right|  <1$}

The general solutions for $a_{c}(t)$, $\chi_{c}(t)$, $P_{a_{c}}(t)$ and
$P_{\chi_{c}}(t)$ in the $\theta<1$\ case\ are
\begin{equation}%
\begin{array}
[c]{c}%
a_{c}(t)=e^{\theta t}\left[  A\sqrt{1-\theta^{2}}\cos\left(  \sqrt
{1-\theta^{2}}t\right)  +B\sqrt{1-\theta^{2}}\sin\left(  \sqrt{1-\theta^{2}%
}t\right)  \right] \\
\\
+e^{-\theta t}\left[  C\sqrt{1-\theta^{2}}\cos\left(  \sqrt{1-\theta^{2}%
}t\right)  +D\sqrt{1-\theta^{2}}\sin\left(  \sqrt{1-\theta^{2}}t\right)
\right]  ,
\end{array}
\label{24}%
\end{equation}

\begin{equation}%
\begin{array}
[c]{c}%
\chi_{c}(t)=e^{\theta t}\left[  A\sqrt{1-\theta^{2}}\cos\left(  \sqrt
{1-\theta^{2}}t\right)  +B\sqrt{1-\theta^{2}}\sin\left(  \sqrt{1-\theta^{2}%
}t\right)  \right] \\
\\
-e^{-\theta t}\left[  C\sqrt{1-\theta^{2}}\cos\left(  \sqrt{1-\theta^{2}%
}t\right)  +D\sqrt{1-\theta^{2}}\sin\left(  \sqrt{1-\theta^{2}}t\right)
\right]  ,
\end{array}
\label{25}%
\end{equation}

\begin{equation}
P_{a_{c}}=2e^{\theta t}\left[  -B\cos\left(  \sqrt{1-\theta^{2}}t\right)
+A\sin\left(  \sqrt{1-\theta^{2}}t\right)  \right]  +2e^{-\theta t}\left[
-D\cos\left(  \sqrt{1-\theta^{2}}t\right)  +C\sin\left(  \sqrt{1-\theta^{2}%
}t\right)  \right]  , \label{25.5}%
\end{equation}

\begin{equation}
P_{\chi_{c}}=2e^{\theta t}\left[  B\cos\left(  \sqrt{1-\theta^{2}}t\right)
-A\sin\left(  \sqrt{1-\theta^{2}}t\right)  \right]  +2e^{-\theta t}\left[
-D\cos\left(  \sqrt{1-\theta^{2}}t\right)  +C\sin\left(  \sqrt{1-\theta^{2}%
}t\right)  \right]  . \label{26}%
\end{equation}
From (\ref{21}), (\ref{24})-(\ref{26}) we can calculate $a_{nc}(t)$ and
$\chi_{nc}(t)$ as
\begin{equation}%
\begin{array}
[c]{c}%
a_{nc}(t)=e^{\theta t}\left\{  \left[  A\sqrt{1-\theta^{2}}-B\theta\right]
\cos\left(  \sqrt{1-\theta^{2}}t\right)  +\left[  B\sqrt{1-\theta^{2}}%
+A\theta\right]  \sin\left(  \sqrt{1-\theta^{2}}t\right)  \right\} \\
\\
+e^{-\theta t}\left\{  \left[  C\sqrt{1-\theta^{2}}+D\theta\right]
\cos\left(  \sqrt{1-\theta^{2}}t\right)  +\left[  D\sqrt{1-\theta^{2}}%
-C\theta\right]  \sin\left(  \sqrt{1-\theta^{2}}t\right)  \right\}  ,
\end{array}
\label{27}%
\end{equation}
\medskip%

\begin{equation}%
\begin{array}
[c]{c}%
\chi_{nc}(t)=e^{\theta t}\left\{  \left[  A\sqrt{1-\theta^{2}}-B\theta\right]
\cos\left(  \sqrt{1-\theta^{2}}t\right)  +\left[  B\sqrt{1-\theta^{2}}%
+A\theta\right]  \sin\left(  \sqrt{1-\theta^{2}}t\right)  \right\} \\
\\
-e^{-\theta t}\left\{  \left[  C\sqrt{1-\theta^{2}}+D\theta\right]
\cos\left(  \sqrt{1-\theta^{2}}t\right)  +\left[  D\sqrt{1-\theta^{2}}%
-C\theta\right]  \sin\left(  \sqrt{1-\theta^{2}}t\right)  \right\}  .
\end{array}
\label{28}%
\end{equation}
The constraint $\mathcal{H}\approx0$ in the present case can be written as
\begin{equation}
\left(  AC+BD\right)  \sqrt{1-\theta^{2}}+\left(  AD-BC\right)  \theta=0.
\label{28.5}%
\end{equation}
In the limit where $\theta\rightarrow0$, equation (\ref{28.5}) is\ reduced to
the equation (\ref{18}). In the same limit, the NC- and C-frame solutions,
given by (\ref{27})-(\ref{28}) and (\ref{24})-(\ref{25}), respectively,
coincide and match with the commutative solutions (\ref{16}) and (\ref{17}).

From (\ref{24}), (\ref{27}) and (\ref{28.5}) it can be seen that, as in the
commutative case, $a_{nc}(t)$ and $a_{c}(t)$\ are unavoidably singular in the
past and in the future. The exponential multiplying factors in both solutions
can model their shape giving rise to bounces at intermediate times, as is
depicted in the thin and thick solid lines in Fig. $1(a)$ for representative
values of $A,B$ and $C$ [$D$ is fixed by (\ref{28.5})].%
%TCIMACRO{\FRAME{ftbpFU}{5.975in}{3.9254in}{0pt}{\Qcb{The typical behavior of
%the scale factor of the noncommutative FRW universe in the NC-frame
%realization (thick lines) in contrast with C-frame realization (thin lines).
%The scale factor of the\ commutative counterpart appears plotted in the dashed
%lines. $(a)$: $\theta=3/4,$ $A=5,$ $B=3$ and $C=6$. $\left(  b\right)
%:\theta=1,$ $A=3,$ $B=2$ and $C=1$. $\left(  c\right)  :\theta=3/2,$ $A=4,$
%$B=3$ and $C=1$. $\left(  d\right)  :\theta=3/2,$ $A=-1.2,$ $B=2$ and $C=-1$%
%.}}{}{fc1.eps}{\special{ language "Scientific Word";  type "GRAPHIC";
%display "PICT";  valid_file "F";  width 5.975in;  height 3.9254in;
%depth 0pt;  original-width 5.9473in;  original-height 3.8977in;
%cropleft "0";  croptop "1";  cropright "1";  cropbottom "0";
%filename 'FC1.eps';file-properties "XNPEU";}} }%
%BeginExpansion
\begin{figure}
[ptb]
\begin{center}
\includegraphics[
height=3.9254in,
width=5.975in
]%
{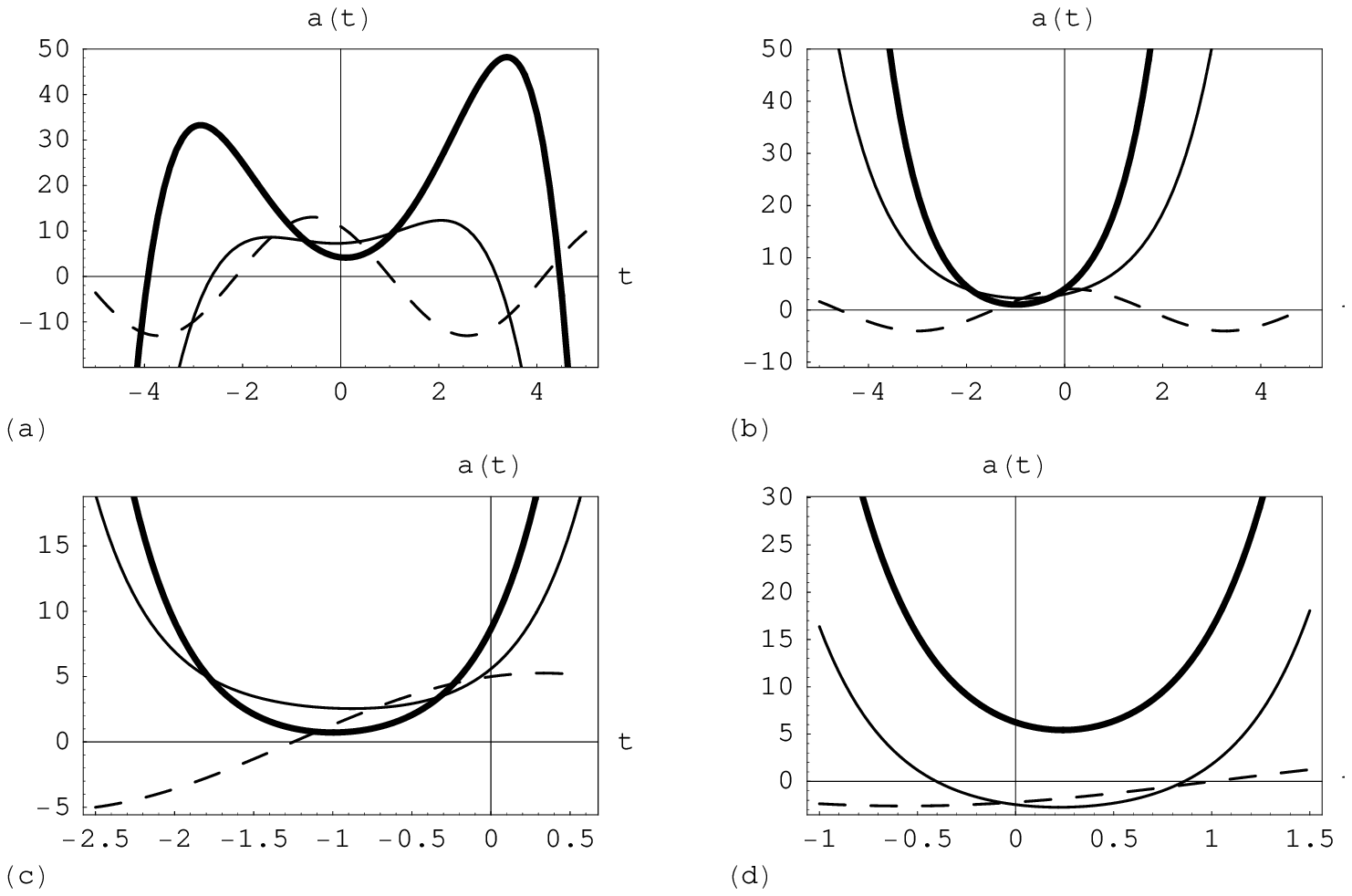}%
\caption{The typical behavior of the scale factor of the noncommutative FRW
universe in the NC-frame realization (thick lines) in contrast with C-frame
realization (thin lines). The scale factor of the\ commutative counterpart
appears plotted in the dashed lines. $(a)$: $\theta=3/4,$ $A=5,$ $B=3$ and
$C=6$. $\left(  b\right)  :\theta=1,$ $A=3,$ $B=2$ and $C=1$. $\left(
c\right)  :\theta=3/2,$ $A=4,$ $B=3$ and $C=1$. $\left(  d\right)
:\theta=3/2,$ $A=-1.2,$ $B=2$ and $C=-1$.}%
\end{center}
\end{figure}
%EndExpansion

\subsubsection{Case $\theta=\pm1$}

When $\theta=\pm1,$ the solutions for $a_{c}(t)$, $\chi_{c}(t)$, $P_{a_{c}%
}(t)$ and $P_{\chi_{c}}(t)$ are
\begin{equation}
a_{c}(t)=A\cosh t+B\sinh t, \label{29}%
\end{equation}

\begin{equation}
\chi_{c}(t)=\pm B\cosh t\pm A\sinh t, \label{30}%
\end{equation}%
\begin{equation}
P_{a_{c}}=2\left(  D+At\right)  \cosh t+2\left(  C+Bt\right)  \sinh t,
\label{31}%
\end{equation}%
\begin{equation}
P_{\chi_{c}}=\mp2\left(  C+Bt\right)  \cosh t\mp2\left(  D+At\right)  \sinh t.
\label{32}%
\end{equation}
As the corresponding $a_{nc}(t)$ and $\chi_{nc}(t),$\ we have
\begin{equation}
a_{nc}(t)=\left(  A+C+Bt\right)  \cosh t+\left(  B+D+At\right)  \sinh t,
\label{33}%
\end{equation}%
\begin{equation}
\chi_{nc}(t)=\pm\left(  B+D+At\right)  \cosh t\pm\left(  A+C+Bt\right)  \sinh
t. \label{34}%
\end{equation}
The constraint $\mathcal{H}\approx0$ in the present case can be written as
\begin{equation}
A^{2}-B^{2}+2\left(  AC-BD\right)  =0. \label{34.5}%
\end{equation}

As it can be seen, the universe solutions in the case $\theta=\pm1$ are
characterized by a qualitative behavior that\ differs from the one
corresponding to the case$\ \left\vert \theta\right\vert <1$. There exist
non-singular bouncing solutions for both $a_{nc}(t)$ and $a_{c}(t)$, as
depicted in Fig. $1(b)$. However, the correspondence between the NC- and
C-frames can be broken for some values of the integration constants.
Non-singular solutions in the NC-frame can correspond to singular solutions in
the C-frame. An interesting example is the case where $A=B=0$ and
$C>\left\vert D\right\vert $. This corresponds to a bouncing universe in the
NC-frame that has no counterpart in the\ C-frame, where the universe is
singular at all times.

\subsubsection{Case $\left\vert \theta\right\vert >1$}

The general solutions for $a_{c}(t)$, $\chi_{c}(t)$, $P_{a_{c}}(t)$ and
$P_{\chi_{c}}(t)$ when $\left\vert \theta\right\vert >1\ $are
\begin{equation}%
\begin{array}
[c]{c}%
a_{c}(t)=e^{\theta t}\left[  A\sqrt{\theta^{2}-1}\cosh\left(  \sqrt{\theta
^{2}-1}t\right)  +B\sqrt{\theta^{2}-1}\sinh\left(  \sqrt{\theta^{2}%
-1}t\right)  \right] \\
\\
+e^{-\theta t}\left[  C\sqrt{\theta^{2}-1}\cosh\left(  \sqrt{1-\theta^{2}%
}t\right)  +D\sqrt{\theta^{2}-1}\sinh\left(  \sqrt{\theta^{2}-1}t\right)
\right]  ,
\end{array}
\label{35}%
\end{equation}

\begin{equation}%
\begin{array}
[c]{c}%
\chi_{c}(t)=e^{\theta t}\left[  A\sqrt{\theta^{2}-1}\cosh\left(  \sqrt
{\theta^{2}-1}t\right)  +B\sqrt{\theta^{2}-1}\sinh\left(  \sqrt{\theta^{2}%
-1}t\right)  \right] \\
\\
-e^{-\theta t}\left[  C\sqrt{\theta^{2}-1}\cosh\left(  \sqrt{\theta^{2}%
-1}t\right)  +D\sqrt{\theta^{2}-1}\sinh\left(  \sqrt{\theta^{2}-1}t\right)
\right]  ,
\end{array}
\label{36}%
\end{equation}

\begin{equation}
P_{a_{c}}=2e^{\theta t}\left[  B\cosh\left(  \sqrt{\theta^{2}-1}t\right)
+A\sinh\left(  \sqrt{\theta^{2}-1}t\right)  \right]  +2e^{-\theta t}\left[
D\cosh\left(  \sqrt{\theta^{2}-1}t\right)  +C\sin\left(  \sqrt{\theta^{2}%
-1}t\right)  \right]  , \label{37}%
\end{equation}

\begin{equation}
P_{\chi_{c}}=-2e^{\theta t}\left[  B\cos\left(  \sqrt{\theta^{2}-1}t\right)
+A\sin\left(  \sqrt{\theta^{2}-1}t\right)  \right]  +2e^{-\theta t}\left[
D\cos\left(  \sqrt{\theta^{2}-1}t\right)  +C\sinh\left(  \sqrt{\theta^{2}%
-1}t\right)  \right]  . \label{37.5}%
\end{equation}
From (\ref{21}), (\ref{24})-(\ref{26}) we can calculate the corresponding
$a_{nc}(t)$ and $\chi_{nc}(t)$ as
\begin{equation}%
\begin{array}
[c]{c}%
a_{nc}(t)=e^{\theta t}\left\{  \left[  A\sqrt{\theta^{2}-1}+B\theta\right]
\cosh\left(  \sqrt{\theta^{2}-1}t\right)  +\left[  B\sqrt{\theta^{2}%
-1}+A\theta\right]  \sinh\left(  \sqrt{\theta^{2}-1}t\right)  \right\} \\
\\
+e^{-\theta t}\left\{  \left[  C\sqrt{\theta^{2}-1}-D\theta\right]
\cosh\left(  \sqrt{\theta^{2}-1}t\right)  +\left[  D\sqrt{\theta^{2}%
-1}-C\theta\right]  \sinh\left(  \sqrt{\theta^{2}-1}t\right)  \right\}  ,
\end{array}
\label{38}%
\end{equation}

\begin{equation}%
\begin{array}
[c]{c}%
\chi_{nc}(t)=e^{\theta t}\left\{  \left[  A\sqrt{\theta^{2}-1}+B\theta\right]
\cosh\left(  \sqrt{\theta^{2}-1}t\right)  +\left[  B\sqrt{\theta^{2}%
-1}+A\theta\right]  \sinh\left(  \sqrt{\theta^{2}-1}t\right)  \right\} \\
\\
-e^{-\theta t}\left\{  \left[  C\sqrt{\theta^{2}-1}-D\theta\right]
\cosh\left(  \sqrt{\theta^{2}-1}t\right)  +\left[  D\sqrt{\theta^{2}%
-1}-C\theta\right]  \sinh\left(  \sqrt{\theta^{2}-1}t\right)  \right\}  .
\end{array}
\label{39}%
\end{equation}
\ When $\left\vert \theta\right\vert >1,$ the constraint $\mathcal{H}\approx0$
is reduced to
\begin{equation}
\left(  BD-AC\right)  \sqrt{\theta^{2}-1}+\left(  AD-BC\right)  \theta=0.
\label{41}%
\end{equation}

As in the case where $\theta=\pm1$,\ there are present non-singular bouncing
solutions in both NC- and C-frames. Fig. $1(c)$ depicts one example. Again we
find that, depending on the values of the integration constants,\ non-singular
universes in the NC-frame can correspond to singular universes in the C-frame
[Fig. $1(d)$].

It is now clear that the $\theta=\pm1$ case\ establishes a division between
qualitatively different ensembles of noncommutative\ universe solutions, since
only for $\left|  \theta\right|  \geq1$ non-singular universes can exist. But
this is not the only interesting property that appears when $\theta=\pm1.$ As
it will be shown later, the solutions in\ this case have a particular behavior
that reveals the capability noncommutativity has to mimic quantum effects
under special conditions.

\section{Minisuperspace Quantization}

Here we present the quantum version of the commutative universe model
discussed in section 2. The FRW universe with conformally coupled scalar field
has already been investigated in the basis of the Wheeler-DeWitt equation in
\cite{17} and \cite{18}, the latter using Bohmian trajectories. However, in
reference \cite{18} there was a restriction to the regime of small scale
parameter, and the wavefunctions considered were different from the ones
studied in this work.

The quantization of minisuperspace model is here carried out by employing the
Dirac formalism (for details see \cite{18}). By making the canonical
replacement $P_{a}=-i\partial/\partial a$ and $P_{\chi}=-i\partial
/\partial\chi$ in (\ref{12}) we obtain, applying the Dirac quantization
procedure,
\begin{equation}
\left[  -\frac{\partial^{2}}{\partial a^{2}}+\frac{\partial^{2}}{\partial
\chi^{2}}+4\left(  a^{2}-\chi^{2}\right)  \right]  \Psi(a,\chi)=0, \label{45}%
\end{equation}
which is the Wheeler-DeWitt equation for the conformally coupled scalar field
model.\footnote{A particular factor ordering is being assumed.} This equation
can be solved by separating\ the $a$ and $\chi$ variables, as it has been done
in the literature (see \cite{18,22.1} and references therein). However, by
making a change of variables there is another route to tackle the
problem\ that is\ interesting by the ensemble of solutions it generates. We
shall present it here, and show, later on, that such a route it is
particularly suitable for application in the noncommutative quantum case.

By making the coordinate change
\begin{equation}
a=\xi\cosh\eta,\text{ \ \ \ }\chi=\xi\sinh\eta,\label{46}%
\end{equation}
we can rewrite (\ref{45}) as
\begin{equation}
\left[  \left(  \frac{\partial^{2}}{\partial\xi^{2}}+\frac{1}{\xi}%
\frac{\partial}{\partial\xi}-\frac{1}{\xi^{2}}\frac{\partial^{2}}{\partial
\eta^{2}}\right)  -4\xi^{2}\right]  \Psi(\xi,\eta)=0.\label{47}%
\end{equation}
By plugging in the \textit{ansatz}
\begin{equation}
\Psi(\xi,\eta)=R\left(  \xi\right)  e^{i\alpha\eta},\label{48}%
\end{equation}
in (\ref{47}) we obtain, after simplification,
\begin{equation}
\frac{\partial^{2}R}{\partial\xi^{2}}+\frac{1}{\xi}\frac{\partial R}%
{\partial\xi}+\left(  \frac{\alpha^{2}}{\xi^{2}}-4\xi^{2}\right)
R=0.\label{49}%
\end{equation}
A solution to (\ref{49}) is
\[
R\left(  \xi\right)  =AK_{i\alpha/2}\left(  \xi^{2}\right)  +BI_{i\alpha
/2}\left(  \xi^{2}\right)  ,
\]
where $K_{\nu}(x)$ and $I_{\nu}(x)$ are Bessel functions of the second kind,
$A$ and $B$ are constants and $\alpha$ is a real number. The solution of the
Wheeler-DeWitt equation (\ref{47}) is therefore
\begin{equation}
\Psi(\xi,\eta)=AK_{i\alpha/2}\left(  \xi^{2}\right)  e^{i\alpha\eta
}+BI_{i\alpha/2}\left(  \xi^{2}\right)  e^{i\alpha\eta}.\label{50}%
\end{equation}
Such a kind of wavefunction also appears, e.g., in the study quantum wormholes
\cite{27} and in quantum cosmology of the Kantowski-Sachs universe
\cite{15,28}. The contribution corresponding to $I_{\nu}(x)$ is usually
discarded because it leads to a solution that is divergent in the classically
forbidden region of the potential. From the point of view of the quantum
trajectories, as it will be clear in the next subsection, there is no
fundamental reason for this solution be discarded. However, since in this work
our main interest is in the influence of noncommutativity in cosmology, rather
than in the foundations of quantum theory, we shall give preference for
wavefunctions that are also\ admissible in interpretations other than the
Bohmian one. We shall\ therefore discard the $I_{\nu}(x)$ contribution and
write the solution of (\ref{47}) as\footnote{Since $\alpha$ is a continuous
index, in the most general case the summation can be replaced by an
integral.}
\begin{equation}
\Psi(\xi,\eta)=\sum_{\alpha}A_{\alpha}K_{i\alpha/2}\left(  \xi^{2}\right)
e^{i\alpha\eta}.\label{51}%
\end{equation}

\subsection{Quantum Trajectory Formalism}

In order establish a framework where all versions of the universe model can be
compared, it is interesting to appeal to a common language. This is provided
by the Bohmian quantum trajectory formalism, which we briefly describe in this
section. For a more detailed account of the subject, see the references given
in the introduction.

In the formulation presented here,\ we shall benefit from ideas proposed in
\cite{20.5}. The wavefunction will be assumed as not a constituent of the
physical system, as originally assumed by Bohm \cite{19}, but as a
representation of it. Quantum information theory tells us that the
wavefunction has a non-physical character \cite{29}. Bohmian quantum physics
should in some way be in accordance with this fact. Actually, the
comprehension of the meaning of the wavefunction as a representation of a
quantum system is crucial to achieving the understanding of quantum mechanics
from any perspective \cite{20.5}. The same point of view seems to be suitable
for adoption in quantum cosmology. We shall briefly comment this aspect when
applying the Bohmian formalism in the description of the noncommutative
version of our universe model.

When dealing with a\ quantum model one must have a clear picture of the
elements of ontology of the theory. By the elements of ontology, we mean what
the theory is essentially about.\footnote{Every physical theory must
necessarily be essentially\ about something, the \textit{primitive ontology
}of the theory \cite{19.3}. In this sense, all the theories are ontological.}
The orthodox quantum theory based on\ the Copenhagen interpretation, e.g., is
about observers that realize measurements.\ In the Bohmian interpretation, on
the other hand, quantum theory is concerned with the physical systems, which
can be particles, waves, strings, etc.\ In this work the object of attention
is the primordial quantum universe, characterized, in the minisuperspace
formalism, by the configuration variables $a$ and $\chi$. Having fixed the
objects of ontology of the theory, we must determine how\ they evolve in time.
This is done with the aid of the wavefunction, whose role is to provide us the
evolution law. The procedure is best illustrated in the context of
non-relativistic quantum mechanics.

Bohmian non-relativistic quantum mechanics is concerned with the behavior of
point particles that move in space describing quantum trajectories.\ An
evolution law is ascribed to them according to the rule%

\begin{equation}
\dot{x}^{i}=\operatorname{Re}\left\{  \frac{1}{m}\frac{\left[  \Psi^{\ast
}\left(  -i\hbar\partial_{i}\right)  \Psi\right]  }{\Psi^{\ast}\Psi}\right\}
=\frac{\nabla S}{m}, \label{60}%
\end{equation}
where $\Psi$ is the wavefunction and $S$ is obtained from the polar
decomposition $\Psi=A\exp(iS)$. As in the orthodox interpretation, the
wavefunction satisfies the Schr\"{o}dinger equation
\begin{equation}
i\hbar\frac{\partial\Psi}{\partial t}=-\frac{\hbar^{2}}{2m}\nabla^{2}%
\Psi+V\Psi. \label{61}%
\end{equation}

Equations (\ref{60})\ and (\ref{61}) specify completely the theory. Without
any other axiom, all phenomena governed by nonrelativistic quantum mechanics,
from spectral lines and quantum interference phenomena to scattering theory,
superconductivity and quantum computation follow from the analysis of the
dynamical system defined by (\ref{60})\ and (\ref{61}) \cite{20.3}. The
expectation value of a physical quantity associated with a Hermitian operator
$\widehat{A}\left(  \hat{x}^{i},\hat{p}^{i}\right)  $ in the standard
formalism can\ computed in the Bohmian formulation by ensemble averaging the
corresponding ``beable''%

\begin{equation}
\mathcal{B}(\widehat{A})=\operatorname{Re}\left\{  \frac{\left[  \Psi^{\ast
}\widehat{A}\left(  \hat{x}^{i},-i\hbar\partial_{i}\right)  \Psi\right]
}{\Psi^{\ast}\Psi}\right\}  =A\left(  x^{i},t\right)  ,\label{61.5}%
\end{equation}
which represents the same quantity when seen from the Bohmian
perspective.\footnote{Holland \cite{20} calls the procedure defined in
equation (\ref{61.5}) as \textquotedblleft taking the local\ expectation
value\textquotedblright\ of the observable $\widehat{A}$. Such a nomenclature
is not adopted here because it is unsuitable to be used in theories where the
object of ontology is an individual system, as in quantum cosmology.} In the
context of non-relativistic quantum mechanics, it can be shown from first
principles that an ensemble of particles obeying the evolution law (\ref{60})
the associated probability density in the configuration space must be given by
$\rho=\left\vert \Psi\right\vert ^{2}$ \cite{20.3}. This is why computing the
ensemble average of $A\left(  x^{i},t\right)  ,$%
\begin{equation}
\int d^{3}x\rho A\left(  x^{i},t\right)  =\int d^{3}x\Psi^{\ast}\widehat
{A}\left(  \hat{x}^{i},-i\hbar\partial_{i}\right)  \Psi=\langle\widehat
{A}\rangle_{t},\label{62}%
\end{equation}
we arrive at the same results of the standard operatorial formalism. Notice
that the law of motion (\ref{60}) can itself\ be obtained from (\ref{61.5}) by
associating $\dot{x}^{i}$ with the \textquotedblleft beable\textquotedblright%
\ corresponding to the velocity operator
\begin{equation}
\dot{x}^{i}=\mathcal{B}\left(  i[\hat{H},\hat{x}^{i}]\right)  =\frac{\nabla
S}{m}.\label{62.5}%
\end{equation}

As a consequence of being an objective theory of point particles describing
trajectories in space, Bohmian quantum mechanics does not give to probability
a privileged role. Instead, as discussed in \cite{20.3}, such a formulation
probability is a derived concept, a decurrent of the law of motion of the
point particles. The\ Bohmian formulation is thus eminently suitable for the
study of individual systems, as the primordial quantum universe. In the
remaining of this section, we will be concerned with the application of the
theory to the commutative quantum universe discussed in the beginning of this
section. In the next sections a similar study will be carried out for the
noncommutative quantum case.

\subsection{Application to Quantum Cosmology}

In this subsection we shall apply the Bohmian formalism to obtain information
about the evolution of the quantum FRW universe with a conformally coupled
scalar field. In the description of quantum cosmology employing quantum
trajectories we shall extend the evolution law\ (\ref{62.5}) to the
minisuperspace variables. In the commutative case the resulting Bohmian
minisuperspace formalism matches with the minisuperspace version of the
Bohmian quantum gravity proposed in \cite{20}, and employed to study the
conformally coupled scalar model in \cite{18}. From (\ref{62.5}) we find, in
the gauge $N=a,$%
\begin{equation}
\dot{a}=\operatorname{Re}\left\{  \frac{\left[  \Psi^{\ast}\left(
-i\partial_{a}/2\right)  \Psi\right]  }{\Psi^{\ast}\Psi}\right\}  =-\frac
{1}{2}\frac{\partial S}{\partial a}, \label{69}%
\end{equation}%
\begin{equation}
\dot{\chi}=\operatorname{Re}\left\{  \frac{\left[  \Psi^{\ast}\left(
-i\partial_{\chi}/2\right)  \Psi\right]  }{\Psi^{\ast}\Psi}\right\}  =\frac
{1}{2}\frac{\partial S}{\partial\chi}. \label{69.5}%
\end{equation}
By changing (\ref{69}) and (\ref{69.5}) into the$\ \left(  \xi,\eta\right)  $
coordinates defined by (\ref{46}) we obtain
\begin{equation}
\frac{d\xi}{dt}=-\frac{1}{2}\frac{\partial S(\xi,\eta)}{\partial\xi
}\text{,\ \ }\ \frac{d\eta}{dt}=\frac{1}{2\xi^{2}}\frac{\partial S(\xi,\eta
)}{\partial\eta}. \label{70}%
\end{equation}

In what follows$^{{}}$ we shall solve the system (\ref{70}) in two examples
where the a universe is characterized by a wavefunction of the type (\ref{51}).

\subsubsection{Case 1}

Let us consider first the example where there is a single Bessel function in
(\ref{51}). In this case the wavefunction is given by
\begin{equation}
\Psi(\xi,\eta)=AK_{i\alpha/2}\left(  \xi^{2}\right)  e^{i\alpha\eta},
\label{74}%
\end{equation}
where $A$ is a constant. Since the Bessel function $K_{i\nu}(x)$ is real for
$\nu$ real and $x>0$,\footnote{This can be verified by looking at the integral
representation (8.432) in page 958 of reference \cite{26.5}.} the phase can be
read directly from the exponential in (\ref{74}): $S=\alpha\eta$. The
equations of motion (\ref{70}) in this state are therefore\ reduced to
\begin{equation}
\frac{d\xi}{dt}=0\text{,\ \ }\ \frac{d\eta}{dt}=\frac{\alpha}{2\xi^{2}},
\label{76}%
\end{equation}
whose solutions are
\begin{equation}
\xi=\xi_{0}\text{, \ \ }\eta=\frac{\alpha}{2\xi_{0}^{2}}t+\eta_{0}. \label{77}%
\end{equation}
The corresponding $a(t)$ and $\chi(t)$ obtained from (\ref{46}) are\ given by
\begin{equation}
a\left(  t\right)  =\xi_{0}\cosh\left(  \frac{\alpha}{2\xi_{0}^{2}}+\eta
_{0}\right)  ,\text{ \ \ \ }\chi\left(  t\right)  =\xi_{0}\sinh\left(
\frac{\alpha}{2\xi_{0}^{2}}t+\eta_{0}\right)  . \label{78}%
\end{equation}

Quantum effects can therefore remove the cosmological singularity, giving rise
to bouncing universes. Under suitable conditions, solution (\ref{78}) can
represent a quantum universe that is indistinguishable from a noncommutative
classical universe. This is seen by noting that it\ can be mapped into the
solution for $a_{c}(t)$ (Eq. \ref{29}) in the case where $\theta=\pm1$ by
identifying $\alpha/2\xi_{0}^{2}=1$, $\xi_{0}=A$ and $\eta_{0}=B=0$, or to the
solution for $a_{nc}(t)$ (Eq. \ref{33}) by identifying $\alpha/2\xi_{0}%
^{2}=1,$ $\xi_{0}=C$ and $\eta_{0}=A=B=D=0$.

In the case where $\alpha=0,$ equation (\ref{78}) describes a static universe
with arbitrary large scale factor, a highly nonclassical behavior.

\subsubsection{Case 2}

Let us now consider a wavefunction that is a superposition of two Bessel
functions in (\ref{51}), that is
\begin{equation}
\Psi(\xi,\eta)=A_{1}K_{i\mu/2}\left(  \xi^{2}\right)  e^{i\mu\eta}%
+A_{2}K_{i\nu/2}\left(  \xi^{2}\right)  e^{i\nu\eta}. \label{80}%
\end{equation}
As the corresponding phase we find
\begin{equation}
S(\xi,\eta)=\arctan\left[  \frac{A_{1}K_{i\mu/2}\left(  \xi^{2}\right)
\sin\left(  \mu\eta\right)  +A_{2}K_{i\nu/2}\left(  \xi^{2}\right)
\sin\left(  \nu\eta\right)  }{A_{1}K_{i\mu/2}\left(  \xi^{2}\right)
\cos\left(  \mu\eta\right)  +A_{2}K_{i\nu/2}\left(  \xi^{2}\right)
\cos\left(  \nu\eta\right)  }\right]  , \label{81}%
\end{equation}
where the $A_{1}$ and $A_{2}$\ were chosen\ as real coefficients. The
equations of motion (\ref{70}) for this state are
\begin{equation}
\frac{d\xi}{dt}=-\frac{A_{1}A_{2}\left[  K_{i\mu/2}^{\prime}K_{i\nu/2}%
-K_{i\mu/2}K_{i\nu/2}^{\prime}\right]  \xi\sin\left[  \left(  \mu-\nu\right)
\eta\right]  }{A_{1}^{2}K_{i\mu/2}^{2}+A_{2}^{2}K_{i\nu/2}^{2}+2A_{1}%
A_{2}K_{i\mu/2}K_{i\nu/2}\cos\left[  \left(  \mu-\nu\right)  \eta\right]
},\hspace{1.8cm} \label{82}%
\end{equation}%
\begin{equation}
\frac{d\eta}{dt}=\frac{1}{2\xi^{2}}\frac{\mu A_{1}^{2}K_{i\mu/2}^{2}+\nu
A_{2}^{2}K_{i\nu/2}^{2}+\left(  \mu+\nu\right)  A_{1}A_{2}K_{i\mu/2}K_{i\nu
/2}\cos\left[  \left(  \mu-\nu\right)  \eta\right]  }{A_{\mu}^{2}K_{i\mu
/2}^{2}+A_{\nu}^{2}K_{i\nu/2}^{2}+2A_{1}A_{2}K_{i\mu/2}K_{i\nu/2}\cos\left[
\left(  \mu-\nu\right)  \eta\right]  }, \label{83}%
\end{equation}
where prime means derivative with respect to the argument. The system
(\ref{82})-(\ref{83}) is set of nonlinear coupled differential equations.
Analytical solutions are difficult to find. Numerical solutions, on the other
hand, can be easily computed\ for $\xi(t)$ and $\eta(t)$. Once the solutions
for $\xi(t)$ and $\eta(t)$ are found, the corresponding $a(t)$ and $\chi(t)$
are determined from (\ref{46}).

The qualitative properties of the solutions of autonomous system such as
(\ref{82})-(\ref{83}) can be determined by analyzing the associated field of
velocities. From the right hand side (RHS0 of (\ref{82})-(\ref{83}) we can see
that the field of velocities has its direction inverted under the substitution
$\mu\rightarrow-\mu,$ $\nu\rightarrow-\nu$. Therefore, to have a qualitative
picture of the associated flow one must be concerned only with the relative
sign of $\mu$ and $\nu$. For simplicity, let us fix $A_{1}=A_{2}=1/\sqrt{2}$,
and, without loss of generality, consider $\mu>0.$ The direction fields
corresponding to the two possible combinations of relative sign between $\mu$
and $\nu$\ are\ depicted in Figs. $2(a)$ and$\ 2(b)$ for representative values
of these constants, where the orbits of some solutions are also plotted. As it
can be seen, the case with positive $\mu$ and negative $\nu$\ favors the
formation of closed orbits, which correspond to non-singular cyclic universes.
The case with positive $\mu$ and $\nu$, on the other hand, tends to privilege
the open orbits.

The presence of the trigonometric functions in the RHS of (\ref{82}%
)-(\ref{83}) is the reason for the repetitive pattern of the direction fields
observed along the $\eta$ direction with period $2\pi/\left\vert \mu
-\nu\right\vert \simeq5.32$ in Figs. $2(a)$ and$\ 2(b)$. The systematic
appearance of closed orbits along the $\eta$ direction in Fig. $2(a)$, and the
possibility of varying their amplitudes with an appropriate choice of the
initial conditions and the constants $\mu$ and $\nu$, show us that the cyclic
universe solutions can present a variable $a_{\min}.$ Another quantity that is
variable is the\ number of $e$-folds between its maximum and minimum size
configurations. This information can be read directly from the logarithmic
plot of $a(t)$ in Fig. $3(a)$, where the solution depicted corresponds to one
of the, closed orbits of Fig. $2(a)$.\ Another interesting non-singular
solution type is depicted in Fig. $3(b)$. This corresponds to one of the open
orbits of Fig. $2(b)$, and\ is an example of an universe that passes by
sequence of infinite bounces starting in the infinite past and never ending.
The sequence is itself enveloped by a larger bounce$.$

A different non-singular\ solution type is present in the case where $\mu
=-\nu$. Since the phase of such a kind of state is $S=0$, the corresponding
universe is necessarily static.%
%TCIMACRO{\FRAME{ftbpFU}{5.8911in}{4.6613in}{0pt}{\Qcb{The field of directions
%and selected orbits corresponding to the Bohmian differential equations for
%the commutative FRW universe in two cases: $\left(  a\right)  :$ $\mu
%=0.6,\nu=-0.58$. Orbits: $\xi_{0}=1.2$, $\eta_{0}=2$ and $\xi_{0}=1.2,\eta
%_{0}=4$. $\left(  b\right)  :$ $\mu=0.6,\nu=1.78.$ Orbits: $\xi_{0}=0.2$,
%$\eta_{0}=3$ and$\ \xi_{0}=1.7$,$\ \eta_{0}=4.5.$}}{}{fc2.eps}%
%maintain-aspect-ratio TRUE;  display "PICT";  valid_file "F";
%width 5.8911in;  height 4.6613in;  depth 0pt;  original-width 5.8643in;
%original-height 4.6337in;  cropleft "0";  croptop "1";  cropright "1";
%cropbottom "0";  filename 'FC2.eps';file-properties "XNPEU";}} }%
%BeginExpansion
\begin{figure}
[ptb]
\begin{center}
\includegraphics[
height=4.6613in,
width=5.8911in
]%
{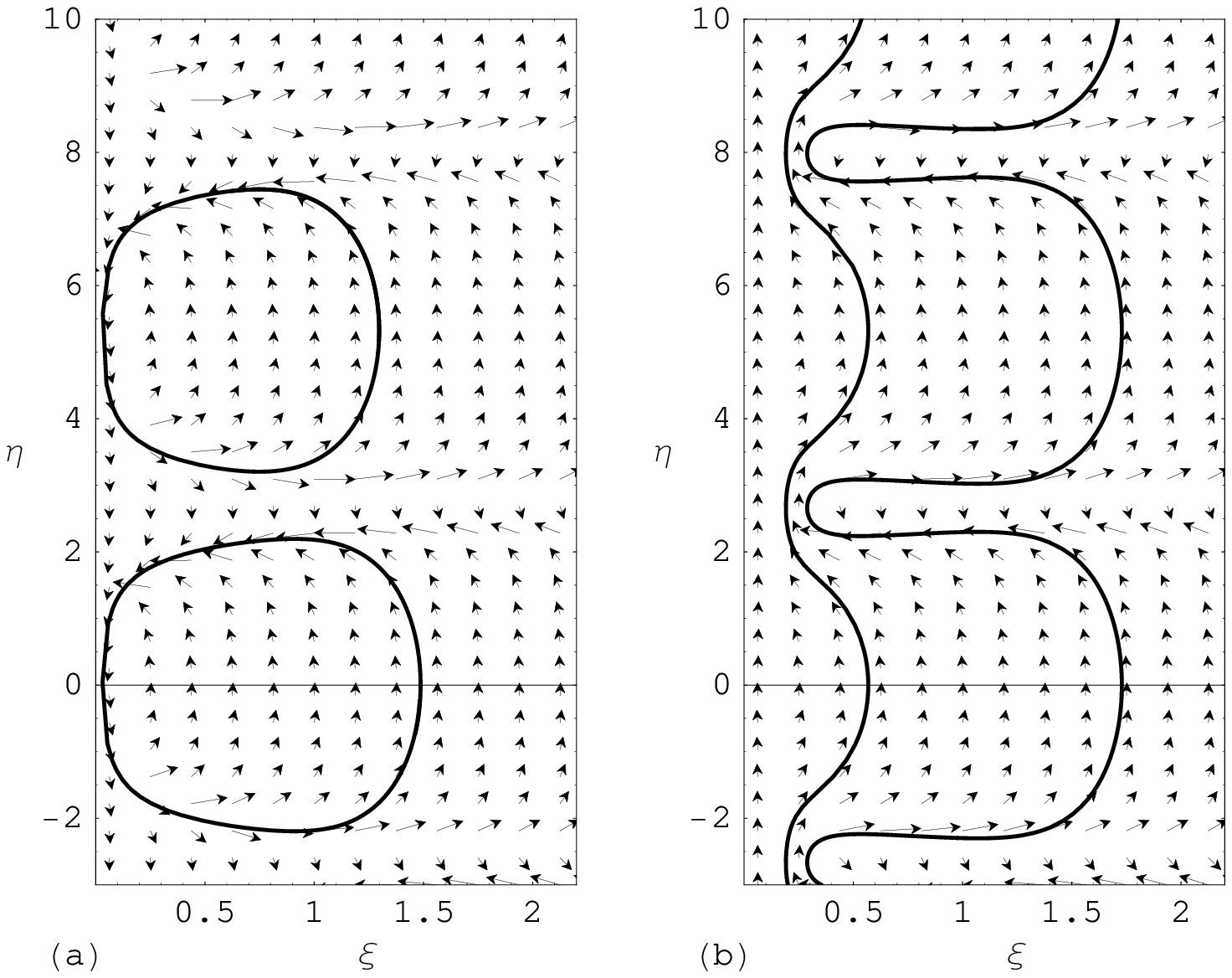}%
\caption{The field of directions and selected orbits corresponding to the
Bohmian differential equations for the commutative FRW universe in two cases:
$\left(  a\right)  :$ $\mu=0.6,\nu=-0.58$. Orbits: $\xi_{0}=1.2$, $\eta_{0}=2$
and $\xi_{0}=1.2,\eta_{0}=4$. $\left(  b\right)  :$ $\mu=0.6,\nu=1.78.$
Orbits: $\xi_{0}=0.2$, $\eta_{0}=3$ and$\ \xi_{0}=1.7$,$\ \eta_{0}=4.5.$}%
\end{center}
\end{figure}
%EndExpansion
%

%TCIMACRO{\FRAME{ftbpFU}{5.8219in}{1.9476in}{0pt}{\Qcb{The typical behavior of
%the scale factor of the commutative FRW universe. $(a)$: $\mu=0.6,$
%$\nu=-0.58$, $\xi_{0}=1.2,$ and$\ \eta_{0}=4$. $\left(  b\right)  :$
%$\mu=0.6,$ $\nu=1.78$, $\xi_{0}=1.7,$ and$\ \eta_{0}=4.5$.}}{}{fc3.eps}%
%maintain-aspect-ratio TRUE;  display "PICT";  valid_file "F";
%width 5.8219in;  height 1.9476in;  depth 0pt;  original-width 5.7943in;
%original-height 1.9199in;  cropleft "0";  croptop "1";  cropright "1";
%cropbottom "0";  filename 'FC3.eps';file-properties "XNPEU";}} }%
%BeginExpansion
\begin{figure}
[ptb]
\begin{center}
\includegraphics[
height=1.9476in,
width=5.8219in
]%
{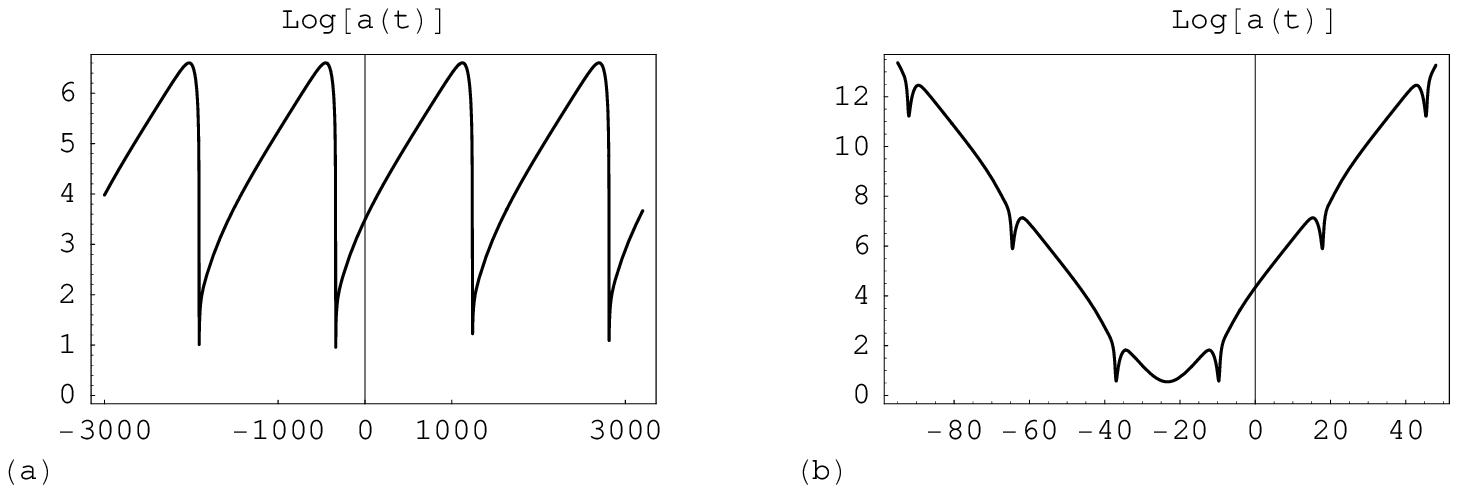}%
\caption{The typical behavior of the scale factor of the commutative FRW
universe. $(a)$: $\mu=0.6,$ $\nu=-0.58$, $\xi_{0}=1.2,$ and$\ \eta_{0}=4$.
$\left(  b\right)  :$ $\mu=0.6,$ $\nu=1.78$, $\xi_{0}=1.7,$ and$\ \eta
_{0}=4.5$.}%
\end{center}
\end{figure}
%EndExpansion

\section{The Noncommutative Quantum Model}

After having studied the individual manifestation of noncommutative and
quantum effects in the conformally coupled scalar field universe, we are now
in time to study the\ combination of them in a unique model. This is achieved
by considering the quantum version of equation (\ref{20}):
\begin{equation}
\lbrack\widehat{a},\widehat{\chi}]=i\theta\text{, \ }[\widehat{a},\widehat
{P}_{a}]=i\text{, \ }[\widehat{\chi},\widehat{P}_{\chi}]=i\text{,\ \ }%
[\widehat{P}_{a},\widehat{P}_{\chi}]=0. \label{100}%
\end{equation}
According to the Weyl quantization procedure \cite{2,3}, the commutation
relation above between the observables $\widehat{a}$ and $\widehat{\chi}$ can
be realized in terms of commutative functions by making use of\ the\ Moyal
star product defined as below%

\begin{equation}
f(a_{c},\chi_{c})\star g(a_{c},\chi_{c})=f(a_{c},\chi_{c})e^{i\frac{\theta}%
{2}\left(  \overleftarrow{\partial}_{a_{c}}\overrightarrow{\partial}_{\chi
_{c}}-\text{ }\overleftarrow{\partial}_{\chi_{c}}\overrightarrow{\partial
}_{a_{c}}\right)  }g(a_{c},\chi_{c}). \label{101}%
\end{equation}

The commutative coordinates $a_{c}$ and $\chi_{c}$ are called Weyl symbols of
the operators $\widehat{a}$ and $\widehat{\chi}$, respectively. A
Wheeler-DeWitt equation the can be adopted for the noncommutative scalar field
universe is%

\begin{equation}
\left[  \widehat{P}_{a_{c}}^{2}-\widehat{P}_{\chi_{c}}^{2}\right]  \Psi
(a_{c},\chi_{c})+4\left(  a_{c}^{2}-\chi_{c}^{2}\right)  \star\Psi(a_{c}%
,\chi_{c})=0, \label{102}%
\end{equation}
which is obtained by Moyal deforming\footnote{For details about this procedure
see \cite{15} and \cite{16}.} (\ref{45}). By using the properties of the Moyal
product, it is possible to write the (\ref{102}) as
\begin{equation}
\left[  \widehat{P}_{a}^{2}-\widehat{P}_{\chi}^{2}\right]  \Psi(a_{c},\chi
_{c})+4\left(  \widehat{a}^{2}-\widehat{\chi}^{2}\right)  \Psi(a_{c},\chi
_{c})=0, \label{103}%
\end{equation}
where%

\begin{equation}
\widehat{a}=\widehat{a}_{c}-\frac{\theta}{2}\widehat{P}_{\chi_{c}}\text{,
\ \ }\widehat{\chi}=\widehat{\chi}\text{$_{c}$}+\frac{\theta}{2}\widehat
{P}_{a_{c}},\text{ \ \ }\widehat{P}_{a_{c}}=\widehat{P}_{a}\text{,
\ \ }\widehat{P}_{\chi_{c}}=\widehat{P}_{\chi}. \label{104}%
\end{equation}
Equation (\ref{104}) is nothing but the operatorial version of\ equation
(\ref{21}). The notation $a_{c}$ and $\chi_{c}$ is now justified. These
symbols match exactly with the canonical variables defined by (\ref{21}). Here
we are in face with the same situation as in the classical case. Two
consistent cosmologies are possible. One considering $\hat{a}_{c}$ and
$\hat{\chi}_{c}$ as the operators associated with the physical metric, and the
other considering $\hat{a}$ and $\hat{\chi}.$ In the application problems we
shall consider the two possibilities.

\subsection{Noncommutative Bohmian formalism}

In order to draw a parallel between the noncommutative quantum universe and
the other three universe types, it is necessary to have a prescription of how
to compute Bohmian trajectories in noncommutative quantum cosmology. The
simplest way to do this is by extending the Bohmian formulation discussed in
section 4 along the same lines proposed in \cite{16}. The procedure consists
in departing from the C-frame and using the \textquotedblleft
beable\textquotedblright\ mapping (\ref{61.5}) to ascribe an evolution law to
the canonical variables. In our time gauge for the noncommutative cosmology,
$N=a_{nc}$ (see section 3), the Hamiltonian (\ref{19}) reduces simply to%
\begin{equation}
h=\left[  -\frac{P_{a_{nc}}^{2}}{4}+\frac{P_{\chi_{nc}}^{2}}{4}-a_{nc}%
^{2}+\chi_{nc}^{2}\right]  .\label{104.5}%
\end{equation}
We can therefore use $h$ to generate time displacements and obtain the
Bohmian\ equations of motion for $a_{c}\left(  t\right)  $ and $\chi
_{c}\left(  t\right)  $ as
\begin{equation}
\frac{da_{c}}{dt}=\mathcal{B}\left(  i[\widehat{h},\widehat{a}_{c}]\right)
=-\frac{1}{2}\left(  1-\theta^{2}\right)  \frac{\partial S}{\partial a_{c}%
}+\theta\chi_{c},\label{105}%
\end{equation}%
\begin{equation}
\frac{d\chi_{c}}{dt}=\mathcal{B}\left(  i[\widehat{h},\widehat{\chi
}\text{$_{c}$}]\right)  =\frac{1}{2}\left(  1-\theta^{2}\right)
\frac{\partial S}{\partial\chi_{c}}+\theta a_{c}.\label{106}%
\end{equation}
The connection between the C- and NC-frame variables is established by
applying the \textquotedblleft beable\textquotedblright\ mapping to the
operatorial equations (\ref{104}), that is, by defining $a\equiv
\mathcal{B}\left(  \widehat{a}\right)  $ and $\chi\equiv\mathcal{B}\left(
\widehat{\chi}\right)  $. Once the trajectories are determined in the C-frame,
one can find their counterparts in the NC-frame by evaluating the variables
$a$ and $\chi$ along the C-frame trajectories,
\begin{equation}
a(t)=\left.  \mathcal{B}\left(  \widehat{a}\right)  \right\vert
_{_{\substack{a_{c}=a_{c}(t)\\\chi_{c}=\chi_{c}(t)}}}=a_{c}(t)-\frac{\theta
}{2}\partial_{\chi_{c}}S\left[  a_{c}(t),\chi_{c}(t)\right]  ,\label{109}%
\end{equation}%
\begin{equation}
\chi(t)=\left.  \mathcal{B}\left(  \widehat{\chi}\right)  \right\vert
_{_{\substack{a_{c}=a_{c}(t)\\\chi_{c}=\chi_{c}(t)}}}=\chi_{c}(t)+\frac
{\theta}{2}\partial_{a_{c}}S\left[  a_{c}(t),\chi_{c}(t)\right]  .\label{110}%
\end{equation}
In what follows we shall\ illustrate the application of the formalism in
noncommutative quantum cosmology.

\section{Application to Noncommutative Quantum Cosmology}

By using the representations $P_{a_{c}}=-i\partial_{a_{c}}$\ and $P_{\chi_{c}%
}=-i\partial_{\chi_{c}}$, we can write the noncommutative Wheeler-DeWitt
equation (\ref{103}) as
\begin{equation}
\left[  \beta\left(  -\frac{\partial^{2}}{\partial a_{c}^{2}}+\frac
{\partial^{2}}{\partial\chi_{c}^{2}}\right)  +4\left(  a_{c}^{2}-\chi_{c}%
^{2}\right)  +4i\theta\left(  \chi_{c}\frac{\partial}{\partial a_{c}}%
+a_{c}\frac{\partial}{\partial\chi_{c}}\right)  \right]  \Psi(a_{c},\chi
_{c})=0, \label{111}%
\end{equation}
where $\beta=1-\theta^{2}$. The separation of variables can be made by
changing to the new set of coordinates
\begin{equation}
a_{c}=\xi\cosh\eta,\text{ \ \ \ }\chi_{c}=\xi\sinh\eta, \label{112}%
\end{equation}
which allow us to rewrite (\ref{111})\ as
\begin{equation}
\left[  \beta\left(  \frac{\partial^{2}}{\partial\xi^{2}}+\frac{1}{\xi}%
\frac{\partial}{\partial\xi}-\frac{1}{\xi^{2}}\frac{\partial^{2}}{\partial
\eta^{2}}\right)  -4i\theta\frac{\partial}{\partial\eta}-4\xi^{2}\right]
\Psi(\xi,\eta)=0. \label{113}%
\end{equation}

The computation of the Bohmian trajectories is rendered easy by expressing the
equations of motion (\ref{105}) and (\ref{106}) in the same hyperbolic
coordinates as the wavefunction. After the change of variables, the Bohmian
equations of motion can be written as
\begin{equation}
\frac{d\xi}{dt}=-\frac{1}{2}\left(  1-\theta^{2}\right)  \frac{\partial
S(\xi,\eta)}{\partial\xi},\text{ \ \ \ }\frac{d\eta}{dt}=\frac{1}{2\xi^{2}%
}\left(  1-\theta^{2}\right)  \frac{\partial S(\xi,\eta)}{\partial\eta}%
+\theta. \label{113.2}%
\end{equation}
The equations (\ref{109}) and (\ref{110}), responsible by the NC-C-frame
correspondence, can be written in the new set of coordinates as%
\begin{equation}
a_{nc}(t)=a_{c}(t)+\frac{\theta}{2}\sinh\eta\text{ }\partial_{\xi}S\left[
\xi(t),\eta(t)\right]  -\frac{\theta}{2}\xi^{-1}\cosh\eta\partial_{\eta
}S\left[  \xi(t),\eta(t)\right]  , \label{113.7}%
\end{equation}%
\begin{equation}
\chi_{nc}(t)=\chi_{c}(t)+\frac{\theta}{2}\cosh\eta\text{ }\partial_{\xi
}S\left[  \xi(t),\eta(t)\right]  -\frac{\theta}{2}\xi^{-1}\sinh\eta
\partial_{\eta}S\left[  \xi(t),\eta(t)\right]  . \label{113.8}%
\end{equation}

Before starting our comparative study by computing Bohmian trajectories
corresponding to specific solutions of (\ref{113}), let us discuss the case of
real wavefunctions.\footnote{This particular case can be of special interest,
since real wavefunctions are priviledged, e.g., by the non-boundary proposal
for the initial conditions of the universe \cite{22.1}.}\ While in the
commutative Bohmian quantum cosmology real wavefunctions always represent
static universes, in the noncommutative Bohmian quantum cosmology they can
represent dynamical universes, a property pointed out in \cite{16} for the
Kantowski-Sachs model. In the FRW with a conformally coupled scalar field
under consideration, equations (\ref{113.2}) tell us that to real
wavefunctions\ correspond always a\ nontrivial and identical dynamics.
Moreover, from (\ref{113.7}) and (\ref{113.8}) we can see that when $S=0$ the
NC- and C-frame realizations are indistinguishable, representing the same
universe. This universe is determined by solving equations (\ref{113.2})\ and
substituting the solutions in (\ref{112}). As a result, we find
\begin{align}
a_{nc}\left(  t\right)   &  =a_{c}\left(  t\right)  =\xi_{0}\cosh\left(
\theta t+\eta_{0}\right)  ,\nonumber\\
& \label{113.9}\\
\chi_{nc}\left(  t\right)   &  =\chi_{c}\left(  t\right)  =\xi_{0}\sinh\left(
\theta t+\eta_{0}\right)  .\nonumber
\end{align}
Real wavefunctions therefore\ always represent non-singular bouncing
universes. Complex wavefunctions, on the other hand, can give rise to a great
variety of dynamics, where the distinction between the frames of physical
realization can be crucial. In the same way as in the classical analog, we can
distinguish three cases: $\left\vert \theta\right\vert <1,\theta=\pm1$ and
$\left\vert \theta\right\vert >1.$

\subsection{Case $\left\vert \theta\right\vert <1$}

In this case, equation (\ref{113}) can be solved by using the \textit{ansatz}
\begin{equation}
\Psi(\xi,\eta)=R\left(  \xi\right)  e^{i\alpha\eta}.\label{114}%
\end{equation}
As a result, we find
\begin{equation}
\frac{\partial^{2}R}{\partial\xi^{2}}+\frac{1}{\xi}\frac{\partial R}%
{\partial\xi}+\left(  \frac{\alpha^{2}}{\xi^{2}}+\frac{4\theta\alpha}{\beta
}-\frac{4\xi^{2}}{\beta}\right)  R=0,\label{115}%
\end{equation}
whose solution is
\begin{equation}
R\left(  \xi\right)  =A_{\alpha}\left(  \frac{\pi}{2}\right)  ^{1/2}\xi
^{-1}W_{\alpha\theta/2\sqrt{\beta},i\alpha/2}\left(  \frac{2\xi^{2}}%
{\sqrt{\beta}}\right)  +B_{\alpha}\left(  \frac{\pi}{2}\right)  ^{1/2}\xi
^{-1}M_{\alpha\theta/2\sqrt{\beta},i\alpha/2}\left(  \frac{2\xi^{2}}%
{\sqrt{\beta}}\right)  ,\label{116}%
\end{equation}
where $W_{u,\nu}(x)$ and $M_{\mu,\nu}(x)$ are Whittaker functions, $A_{\alpha
}$ and $B_{\alpha}$ are constants and $\alpha$ is a real number. The piece
corresponding to the $M_{\mu,\nu}(x)$ contribution leads to a\ divergent
wavefunction in the classically forbidden region. We shall therefore discard
its contribution in a similar way as in the commutative model discussed
before. We can thus write the solution of (\ref{113}) as%

\begin{equation}
\Psi(\xi,\eta)=\sum_{\alpha}\left[  A_{\alpha}\left(  \frac{\pi}{2}\right)
^{1/2}\xi^{-1}W_{\alpha\theta/2\sqrt{\beta},i\alpha/2}\left(  \frac{2\xi^{2}%
}{\sqrt{\beta}}\right)  e^{i\alpha\eta}\right]  . \label{117}%
\end{equation}
In the limit were $\theta=0,$ the Whittaker functions $W_{\mu,i\nu}(x)$ are
reduced to the Bessel functions$\ K_{i\nu}$ through the relation,
\begin{equation}
\xi^{-1}W_{0,i\alpha/2}\left(  2\xi^{2}\right)  =\left(  \frac{2}{\pi}\right)
^{1/2}K_{i\alpha/2}\left(  \xi^{2}\right)  , \label{118}%
\end{equation}
and therefore the wavefunction (\ref{117}) matches with the commutative
wavefunction (\ref{51}).

In the sequel we present two examples of application of the Bohmian formalism
in the investigation of the properties of the wavefunctions.

\subsubsection{Example 1}

The wavefunction is of the type
\begin{equation}
\Psi(\xi,\eta)=A\xi^{-1}W_{\alpha\theta/2\sqrt{\beta},i\alpha/2}\left(
\frac{2\xi^{2}}{\sqrt{\beta}}\right)  e^{i\alpha\eta}, \label{131}%
\end{equation}
where $A$ is a constant. Since the Whittaker function $W_{\mu,i\nu}(x)$ is
real for $\mu\ $and$\ \nu$ real and $x>0$,\footnote{This can be verified by
looking at the integral representation (9.223) in page 1060 of reference
\cite{26.5}.} the phase can be read directly from the exponential:
$S=\alpha\eta.$ The equations of motion for $\xi$ and $\eta$ in\ this state
are%
\begin{equation}
\frac{d\xi}{dt}=0\text{,\ \ }\ \frac{d\eta}{dt}=\frac{\alpha\beta}{2\xi^{2}%
}+\theta. \label{132}%
\end{equation}
\ As the solutions for $a_{c}(t)$ and $\chi_{c}(t),$ we have
\begin{equation}
a_{c}\left(  t\right)  =a_{c_{0}}\cosh\left(  \kappa t\right)  ,\text{
\ \ \ }\chi_{c}\left(  t\right)  =a_{c_{0}}\sinh\left(  \kappa t\right)  ,
\label{134}%
\end{equation}
where $\kappa=$ $\alpha\beta/2a_{c_{0}}^{2}+\theta$ and$\ \eta_{0}$ was
absorbed by redefining the origin of time.

From (\ref{113.7}) and (\ref{113.8}) we can write
\begin{equation}
a_{nc}(t)=a_{nc_{0}}\cosh\left(  \kappa t\right)  ,\text{ \ \ \ }\chi
_{nc}(t)=a_{nc_{0}}\sinh\left(  \kappa t\right)  , \label{137}%
\end{equation}
where $a_{nc_{0}}=a_{c_{0}}-\alpha\theta/2a_{c_{0}}.$ Again we found bouncing
solutions that can be mapped into the classical solution of the same type with
a suitable identification of the integration constants. But this is not the
only interesting property exhibited in this case. From (\ref{137}) we can see
that\ to each universe in the NC frame there corresponds at least one universe
in the C-frame, as it is shown in Fig. $4$.\ For $\alpha\theta$ positive, the
correspondence is one to one and exists only for $a_{c_{0}}>\sqrt{\left\vert
\alpha\theta\right\vert /2}$. Smaller values of $a_{c_{0}}$ would imply a
singular universe in the NC-frame.

For negative values of $\alpha\theta$, on the other hand, the correspondence
is defined for all values of $a_{c_{0}}$. To each universe in the NC-frame
there correspond two universes in the C-frame. An exception occurs for
$a_{c_{0}}=\sqrt{\alpha\theta/2}$, where the curve $a_{nc_{0}}\times a_{c_{0}%
}$ achieves its minimum and the\ correspondence is one-to-one. This value of
$a_{c_{0}}$\ marks the division between the two regimes that govern the
NC-C-frame\ correspondence: large $a_{nc_{0}}$ and small $a_{c_{0}}$ and of
large $a_{nc_{0}}$ and $a_{c_{0}}$. A similar behavior was previously found in
\cite{22.3} in non-relativistic Bohmian quantum mechanics when studying the
harmonic oscillator. The capability of the noncommutativity effects to promote
the interplay between large and small scale distances was interpreted in that
reference as manifestation of a sort of \textquotedblleft UR-UV
mixing\textquotedblright\ in the oscillator orbits.%
%TCIMACRO{\FRAME{ftbpFU}{4.4287in}{4.3042in}{0pt}{\Qcb{The minimum value of the
%scale factor of the noncommutative quantum FRW universe with $\left\vert
%\theta\right\vert <1$ in the NC-frame $y=a_{nc_{0}}$ as a function of its
%value in the C-frame $x=a_{c_{0}}.$ The thick line refers to the case
%where$\ \alpha\theta<0,$ and the\ thin line to that where $\alpha\theta>0$.
%The values adopted for $\alpha\theta$ are$\ \alpha\theta=-1$ and $\alpha
%\theta=1$. The diagonal line is also plotted.}}{}{fc4.eps}%
%maintain-aspect-ratio TRUE;  display "PICT";  valid_file "F";
%width 4.4287in;  height 4.3042in;  depth 0pt;  original-width 4.401in;
%original-height 4.2756in;  cropleft "0";  croptop "1";  cropright "1";
%cropbottom "0";  filename 'FC4.eps';file-properties "XNPEU";}} }%
%BeginExpansion
\begin{figure}
[ptb]
\begin{center}
\includegraphics[
height=4.3042in,
width=4.4287in
]%
{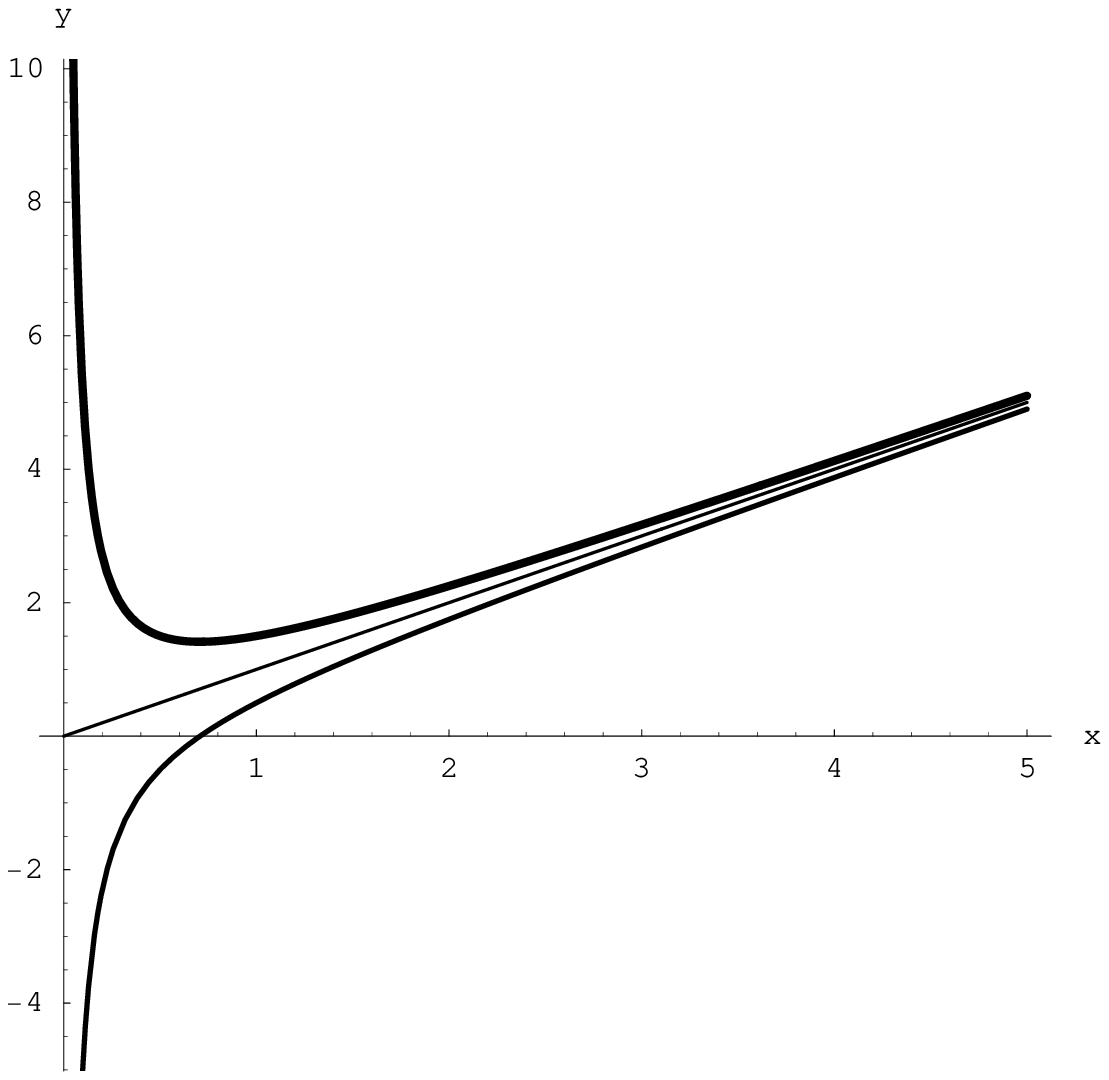}%
\caption{The minimum value of the scale factor of the noncommutative quantum
FRW universe with $\left\vert \theta\right\vert <1$ in the NC-frame
$y=a_{nc_{0}}$ as a function of its value in the C-frame $x=a_{c_{0}}.$ The
thick line refers to the case where$\ \alpha\theta<0,$ and the\ thin line to
that where $\alpha\theta>0$. The values adopted for $\alpha\theta$
are$\ \alpha\theta=-1$ and $\alpha\theta=1$. The diagonal line is also
plotted.}%
\end{center}
\end{figure}
%EndExpansion

\bigskip\bigskip\bigskip

\subsubsection{Example 2}

The wavefunction function (\ref{117})\ is a sum of two functions,
\begin{equation}
\Psi(\xi,\eta)=A_{1}\xi^{-1}W_{\mu\theta/2\sqrt{\beta},i\mu/2}\left(
\frac{2\xi^{2}}{\sqrt{\beta}}\right)  e^{i\mu\eta}+A_{2}\xi^{-1}W_{\nu
\theta/2\sqrt{\beta},i\nu/2}\left(  \frac{2\xi^{2}}{\sqrt{\beta}}\right)
e^{i\nu\eta}. \label{151}%
\end{equation}
The corresponding phase is
\begin{equation}
S(\xi,\eta)=\arctan\left[  \frac{A_{1}W_{\mu\theta/2\sqrt{\beta},i\mu/2}%
\sin\left(  \mu\eta\right)  +A_{2}W_{\nu\theta/2\sqrt{\beta},i\nu/2}%
\sin\left(  \nu\eta\right)  }{A_{1}W_{\mu\theta/2\sqrt{\beta},i\mu/2}%
\cos\left(  \mu\eta\right)  +A_{2}W_{\nu\theta/2\sqrt{\beta},i\nu/2}%
\cos\left(  \nu\eta\right)  }\right]  , \label{152}%
\end{equation}
where, as in the commutative case, $A_{1}$ and $A_{2}$\ were chosen\ as real
coefficients. The equations of motion (\ref{113.2}) in this state are
\begin{equation}
\frac{d\xi}{dt}=-2\xi\sqrt{\beta}\frac{A_{1}A_{2}\left[  W_{\mu\theta
/2\sqrt{\beta},i\mu/2}^{\prime}W_{\nu\theta/2\sqrt{\beta},i\nu/2}-W_{\mu
\theta/2\sqrt{\beta},i\mu/2}W_{\nu\theta/2\sqrt{\beta},i\nu/2}^{\prime
}\right]  \sin\left[  \left(  \mu-\nu\right)  \eta\right]  }{A_{1}^{2}%
W_{\mu\theta/2\sqrt{\beta},i\mu/2}^{2}+A_{2}^{2}W_{\nu\theta/2\sqrt{\beta
},i\nu/2}^{2}+2A_{1}A_{2}W_{\mu\theta/2\sqrt{\beta},i\mu/2}W_{\nu\theta
/2\sqrt{\beta},i\nu/2}\cos\left[  \left(  \mu-\nu\right)  \eta\right]  },
\label{153}%
\end{equation}%
\begin{equation}
\frac{d\eta}{dt}=\frac{\beta}{2\xi^{2}}\frac{\mu A_{1}^{2}W_{\mu\theta
/2\sqrt{\beta},i\mu/2}^{2}+\nu A_{2}^{2}W_{\nu\theta/2\sqrt{\beta},i\nu/2}%
^{2}+A_{1}A_{2}\left(  \mu+\nu\right)  W_{\mu\theta/2\sqrt{\beta},i\mu
/2}W_{\nu\theta/2\sqrt{\beta},i\nu/2}\cos\left[  \left(  \mu-\nu\right)
\eta\right]  }{A_{1}^{2}W_{\mu\theta/2\sqrt{\beta},i\mu/2}^{2}+A_{2}^{2}%
W_{\nu\theta/2\sqrt{\beta},i\nu/2}^{2}+2A_{1}A_{2}W_{\mu\theta/2\sqrt{\beta
},i\mu/2}W_{\nu\theta/2\sqrt{\beta},i\nu/2}\cos\left[  \left(  \mu-\nu\right)
\eta\right]  }+\theta, \label{154}%
\end{equation}
where prime means derivative with respect to the argument. As in the
commutative counterpart, we have an autonomous set of non-linear coupled
differential equations to solve. In order to render easy the comparison with
that case, let us fix $A_{1}=A_{2}=1/\sqrt{2}$.\ Again its is possible to find
bouncing solutions, as it\ is shown in Fig $5(a)$,\ where the splitting
between NC- and C-frame evolutions is quantitatively irrelevant. The effect of
noncommutativity in that case is manifest through the suppression of the
sequence of bounces that appears in the commutative counterpart and\ by
shifting the time where the universe achieves its minimum size [see Fig.
$3(b)$].

The case where $\mu=-\nu$, which in the commutative counterpart corresponds to
a static universe, here has a non-trivial dynamics. One example is presented
in Fig. $5(b)$, where a cyclic solution similar to that of Fig. $3(a)$ is
depicted. The splitting between NC- and C-frame evolutions in this case is
again quantitatively irrelevant.

\subsection{Case $\theta=\pm1$}

As in the classical counterpart, the case $\theta=\pm1$ is marked by a
peculiar behavior. Equation (\ref{113})\ in this case\ is reduced to the first
order partial differential equation
\begin{equation}
\left[  i\frac{\partial}{\partial\eta}\pm\xi^{2}\right]  \Psi(\xi,\eta)=0,
\label{160}%
\end{equation}
whose general solution is
\begin{equation}
\Psi(\xi,\eta)=R\left(  \xi\right)  e^{\pm i\xi^{2}\eta}, \label{161}%
\end{equation}
where $R\left(  \xi\right)  $ is any differentiable function of $\xi.$ The
equations of motion (\ref{113.2}) in this case are reduced to
\begin{equation}
\frac{d\xi}{dt}=0\text{,\ \ }\ \frac{d\eta}{dt}=\pm1, \label{162}%
\end{equation}
whose solutions are
\begin{equation}
\xi=\xi_{0}\text{, \ \ }\eta=\pm t+\eta_{0}. \label{163}%
\end{equation}
As the solutions for $a_{c}(t)$ and $\chi_{c}(t),$ we have
\begin{equation}
a_{c}\left(  t\right)  =\xi_{0}\cosh\left(  \pm t+\eta_{0}\right)
,\text{\ \ }\chi_{c}\left(  t\right)  =\xi_{0}\sinh\left(  \pm t+\eta
_{0}\right)  , \label{164}%
\end{equation}
while the corresponding $a(t)$ and $\chi(t)$ are
\begin{equation}
a_{nc}\left(  t\right)  =\frac{1}{2}\xi_{0}\cosh\left(  \pm t+\eta_{0}\right)
+\xi_{0}\left(  \pm t+\eta_{0}\right)  \sinh\left(  \pm t+\eta_{0}\right)  ,
\label{165}%
\end{equation}%
\begin{equation}
\chi_{nc}\left(  t\right)  =\frac{1}{2}\xi_{0}\sinh\left(  \pm t+\eta
_{0}\right)  +\xi_{0}\left(  \pm t+\eta_{0}\right)  \cosh\left(  \pm
t+\eta_{0}\right)  . \label{166}%
\end{equation}

In both\ NC- and C-frame realizations of noncommutativity we find bounce
solutions whenever $\xi_{0}>0$, while for $\xi_{0}\leq0$ the universe is
necessarily singular [Fig. $5(c)$]. The novelty here is that these universe
solutions are the most general ones available. No matter what is the
wavefunction, the fate of the universe is determined uniquely by $\xi_{0}$. At
first sight it could seem strange that the wavefunction cannot have any
influence on the fate of the universe, independent of its functional form.
However, if we realize that the information provided by the wavefunction is
about the universe evolution law, we find that the wavefunction is playing its
hole providing us equations (\ref{162}) in the same way as in all the other
cases previously discussed. Since the kinetic term is quenched by
noncommutativity effects, what we found in this case is exactly what one would
expect to find: a poor and highly constrained dynamics, similar to the one
that appears when a magnetic field projects a system onto its lowest Landau
level (see \cite{22.2} and ref. therein).

\subsection{Case $\left\vert \theta\right\vert >1$}

In this last case the most general wavefunction that satisfies (\ref{111}) can
be written as
\begin{equation}
\Psi(\xi,\eta)=\sum_{\alpha}\left[  A_{\alpha}\xi^{-1}W_{i\alpha\theta
/2\sqrt{\left\vert \beta\right\vert },i\alpha/2}\left(  \frac{2i\xi^{2}}%
{\sqrt{\left\vert \beta\right\vert }}\right)  e^{i\alpha\eta}+B_{\alpha}%
\xi^{-1}M_{i\alpha\theta/2\sqrt{\left\vert \beta\right\vert },i\alpha
/2}\left(  \frac{2i\xi^{2}}{\sqrt{\left\vert \beta\right\vert }}\right)
e^{i\alpha\eta}\right]  . \label{180}%
\end{equation}
Contrary to the previous cases, the contribution corresponding to
$M_{i\alpha\theta/2\left\vert \beta\right\vert ,i\alpha/2}$ is not divergent
in the classically forbidden region. Moreover, each of the Whittaker functions
$W_{\mu,\nu}(x)$ and $M_{\mu,\nu}(x)$ in this case is complex, and therefore
can give rise to a dynamics that differs from the ones of the examples
previously discussed. For simplicity, we shall consider only\ the example
where
\begin{equation}
\Psi(\xi,\eta)=AW_{i\alpha\theta/2\sqrt{\left\vert \beta\right\vert }%
,i\alpha/2}\left(  \frac{2i\xi^{2}}{\sqrt{\left\vert \beta\right\vert }%
}\right)  e^{i\alpha\eta}. \label{181}%
\end{equation}
For this wavefunction, the equations of motion (\ref{113.2})\ for $\xi(t)$ and
$\eta(t)$ can be shown to be%

\begin{equation}
\frac{d\xi}{dt}=\sqrt{\left\vert \beta\right\vert }\xi-\frac{\sqrt{\left\vert
\beta\right\vert }\alpha\theta}{2\xi}+\frac{\beta}{\xi}\operatorname{Im}%
\left[  \frac{W_{i\alpha\theta/2\sqrt{\left\vert \beta\right\vert }%
+1,i\alpha/2}\left(  2i\xi^{2}/\sqrt{\left\vert \beta\right\vert }\right)
}{W_{i\alpha\theta/2\sqrt{\left\vert \beta\right\vert },i\alpha/2}\left(
2i\xi^{2}/\sqrt{\left\vert \beta\right\vert }\right)  }\right]  , \label{182}%
\end{equation}%
\begin{equation}
\frac{d\eta}{dt}=\frac{\alpha\beta}{2\xi^{2}}+\theta.\hspace*{8cm} \label{183}%
\end{equation}
The equations (\ref{182}) and (\ref{183}) can be solved analytically in the
limit of large $\xi$, where the contribution coming from the term containing
the Whittaker functions in the right hand side of (\ref{182}) can be
approximated by $-2\sqrt{\left\vert \beta\right\vert }\xi.$ In this
regime,\ (\ref{182}) can be simplified to
\begin{equation}
\frac{d\xi}{dt}=-\sqrt{\left\vert \beta\right\vert }\xi. \label{184}%
\end{equation}
The solutions of (\ref{183}) and (\ref{184}) are
\begin{equation}
\xi(t)=\xi_{0}e^{-\sqrt{\left\vert \beta\right\vert }t},\hspace{3.6cm}
\label{185}%
\end{equation}%
\begin{equation}
\eta(t)=\frac{\alpha\sqrt{\left\vert \beta\right\vert }}{4\xi_{0}^{2}}\left(
1-e^{2\sqrt{\left\vert \beta\right\vert }t}\right)  +\theta t+\eta_{0}.
\label{186}%
\end{equation}
As the expressions for $a_{c}(t)$ and $\chi_{c}(t)$ we have
\begin{equation}
a_{c}(t)=a_{c_{0}}e^{-\sqrt{\left\vert \beta\right\vert }t}\cosh\left[
\frac{\alpha\sqrt{\left\vert \beta\right\vert }}{4a_{c_{0}}^{2}}\left(
1-e^{2\sqrt{\left\vert \beta\right\vert }t}\right)  +\theta t\right]  ,
\label{187}%
\end{equation}%
\begin{equation}
\chi_{c}(t)=a_{c_{0}}e^{-\sqrt{\left\vert \beta\right\vert }t}\sinh\left[
\frac{\alpha\sqrt{\left\vert \beta\right\vert }}{4a_{c_{0}}^{2}}\left(
1-e^{2\sqrt{\left\vert \beta\right\vert }t}\right)  +\theta t\right]  ,
\label{188}%
\end{equation}
where $\eta_{0}$ was absorbed by redefining the origin of time. The
corresponding $a_{nc}(t)$ and $\chi_{nc}(t)$ obtained from (\ref{113.7}) and
(\ref{113.8}) are
\begin{align}
a_{nc}(t)  &  =\left(  a_{c_{0}}-\frac{\alpha\theta}{2a_{c_{0}}}%
e^{2\sqrt{\left\vert \beta\right\vert }t}\right)  e^{-\sqrt{\left\vert
\beta\right\vert }t}\cosh\left[  \frac{\alpha\sqrt{\left\vert \beta\right\vert
}}{4a_{c_{0}}^{2}}\left(  1-e^{2\sqrt{\left\vert \beta\right\vert }t}\right)
+\theta t\right] \nonumber\\
& \label{190}\\
&  -\frac{\theta a_{c_{0}}}{\sqrt{\left\vert \beta\right\vert }}%
e^{-\sqrt{\left\vert \beta\right\vert }t}\sinh\left[  \frac{\alpha
\sqrt{\left\vert \beta\right\vert }}{4a_{c_{0}}^{2}}\left(  1-e^{2\sqrt
{\left\vert \beta\right\vert }t}\right)  +\theta t\right]  ,\nonumber
\end{align}%
\begin{align}
\chi_{nc}(t)  &  =\left(  a_{c_{0}}-\frac{\alpha\theta}{2a_{c_{0}}}%
e^{2\sqrt{\left\vert \beta\right\vert }t}\right)  e^{-\sqrt{\left\vert
\beta\right\vert }t}\sinh\left[  \frac{\alpha\sqrt{\left\vert \beta\right\vert
}}{4a_{c_{0}}^{2}}\left(  1-e^{2\sqrt{\left\vert \beta\right\vert }t}\right)
+\theta t\right] \nonumber\\
& \label{191}\\
&  -\frac{\theta a_{c_{0}}}{\sqrt{\left\vert \beta\right\vert }}%
e^{-\sqrt{\left\vert \beta\right\vert }t}\cosh\left[  \frac{\alpha
\sqrt{\left\vert \beta\right\vert }}{4a_{c_{0}}^{2}}\left(  1-e^{2\sqrt
{\left\vert \beta\right\vert }t}\right)  +\theta t\right]  .\nonumber
\end{align}

From (\ref{185})\ we can see that\ the physical meaning of the approximation
assumed is that of early times. Figure $5(d)$ depicts the scale factors
$a_{nc}(t)$\ and $a_{c}(t)$\ in an interval where the approximation proposed
is accurate. Depending on the values of $\theta,\alpha,$ and $a_{c_{0}}$, the
deviation in the behavior of them can be very large. The situation here is
more or less\ similar to that of case 1 of section 6. We shall therefore not
enter the details. Let us consider, for example, the case where $\alpha
\theta<0.$ For any instant of time $t=T$ where the approximations
(\ref{187})-(\ref{191}) are valid, we can see that to\ each universe in the
NC-frame there correspond two universes in the C-frame: one with $a_{c_{0}}%
$\ large\ and the other with $a_{c_{0}}$ small. The graph of $y=a_{c_{0}%
}-\alpha\theta\exp[2\sqrt{\left\vert \beta\right\vert }T]/2a_{c_{0}}$ [which
in (\ref{190}) has a role similar to that of $a_{nc_{0}}$ in case 1, section
6] is identical in shape to that of Fig. 4.%

%TCIMACRO{\FRAME{ftbhFU}{5.8219in}{4.1079in}{0pt}{\Qcb{Selected plots of the
%scale factor of the FRW universe in the NC-frame realization (thick lines) in
%contrast with C-frame realization (thin lines). $\left(  a\right)
%:\theta=-0.9,$ $\mu=0.6,$ $\nu=1.78$, $\xi_{0}=1.7,$ and$\ \eta_{0}=4.5$.
%$\left(  b\right)  :\theta=0.1,$ $\mu=0.3,$ $\nu=-0.3$, $\xi_{0}=1,$
%and$\ \eta_{0}=10.$ $\left(  c\right)  :\theta=1,$ $\xi_{0}=5,$ and$\ \eta
%_{0}=0.$ $\left(  d\right)  :\theta=-1.5,$ $\alpha=1.5$, $a_{0}=2,$
%and$\ \eta_{0}=0.$}}{}{fc5.eps}{\special{ language "Scientific Word";
%type "GRAPHIC";  display "PICT";  valid_file "F";  width 5.8219in;
%height 4.1079in;  depth 0pt;  original-width 5.7943in;
%original-height 4.0802in;  cropleft "0";  croptop "1";  cropright "1";
%cropbottom "0";  filename 'FC5.eps';file-properties "XNPEU";}} }%
%BeginExpansion
\begin{figure}
[tbh]
\begin{center}
\includegraphics[
height=4.1079in,
width=5.8219in
]%
{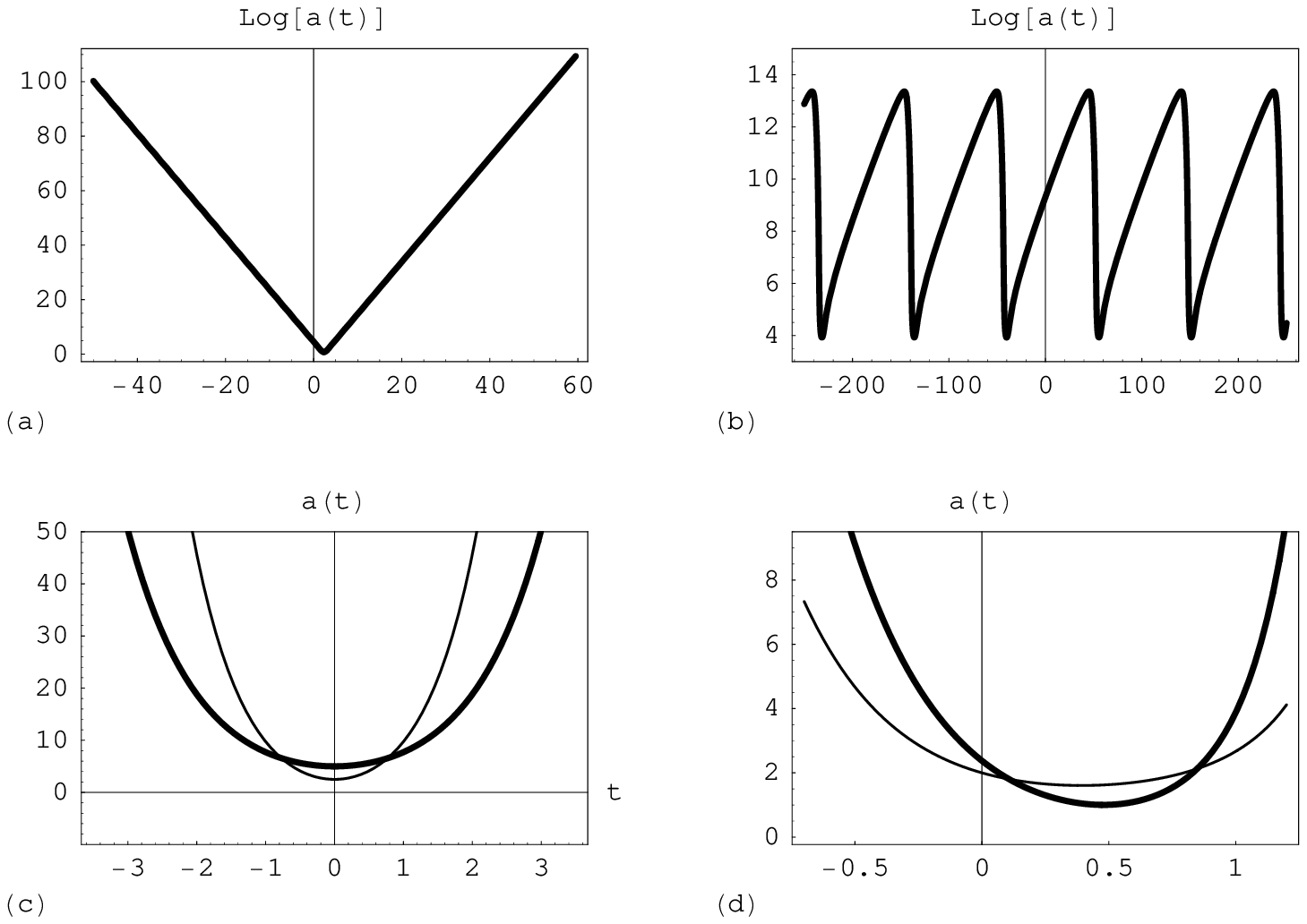}%
\caption{Selected plots of the scale factor of the FRW universe in the
NC-frame realization (thick lines) in contrast with C-frame realization (thin
lines). $\left(  a\right)  :\theta=-0.9,$ $\mu=0.6,$ $\nu=1.78$, $\xi
_{0}=1.7,$ and$\ \eta_{0}=4.5$. $\left(  b\right)  :\theta=0.1,$ $\mu=0.3,$
$\nu=-0.3$, $\xi_{0}=1,$ and$\ \eta_{0}=10.$ $\left(  c\right)  :\theta=1,$
$\xi_{0}=5,$ and$\ \eta_{0}=0.$ $\left(  d\right)  :\theta=-1.5,$ $\alpha
=1.5$, $a_{0}=2,$ and$\ \eta_{0}=0.$}%
\end{center}
\end{figure}
%EndExpansion

\section{Discussion}

In this work, we carried out an investigation into the role of noncommutative
geometry in the cosmological scenario by introducing a noncommutative
deformation in the algebra of the minisuperspace variables along the same
lines proposed in \cite{15} and followed in \cite{16}. As a cosmological model
to carry out such an investigation, we chose a Friedman-Robertson-Walker
universe with conformally coupled scalar field. A parallel was drawn between
the realizations of noncommutativity in two possible frameworks for physical
interpretation: the C-frame, where it manifest through $\theta$-dependent
terms the Hamiltonian, and in the NC-frame, where it is manifest directly in
the universe degrees of freedom.

The influence of noncommutativity in the universe evolution and its capability
to remove cosmological singularities was investigated by means of a
comparative study of the FRW model in four different versions: classical
commutative, classical noncommutative, quantum commutative and quantum
noncommutative. The confrontation between the classical and quantum versions
was rendered easy by the Bohmian interpretation of quantum theory, which
provided a common language for comparison through the quantum trajectory
formalism. An extension of the Bohmian formulation to comprise noncommutative
effects was previously proposed for the Kantowski-Sachs model in \cite{16}. In
our comparative study we have dealt with the noncommutative quantum model
along the same lines. The \textquotedblleft beable\textquotedblright\ mapping
commonly employed in Bohmian quantum mechanics was extended to noncommutative
quantum cosmology. In the commutative context, our formulation is reduced to
the one proposed by Holland \cite{20} in the minisuperspace approximation.

In the classical context, the main result of our investigation is
that,\ contrary to the noncommutative Kantowski-Sachs model, for $\theta$
sufficiently large the noncommutative FRW can be non-singular. When $\left|
\theta\right|  \geq1$, noncommutativity can give rise to bouncing universes in
the NC- and C-frame realizations. The $\theta=\pm1$ case\ is of particular
interest since it reveals the capability noncommutativity has to mimic quantum
effects under special conditions. The bouncing solutions that appear in both
NC- and C-frame realizations [Eqs. (\ref{29}) and (\ref{30})] can be mapped
into the commutative quantum solution (\ref{78}) with an appropriate
identification of the integration constants. Therefore noncommutative
classical quantum universe can be indistinguishable from a commutative quantum
universe. Similar correspondences involving the noncommutative classical
universe and other bouncing solutions of noncommutative quantum cosmology were
described in section 6.

While in the classical context non-singular universe solutions can exist only
in the noncommutative universe model and for $\left\vert \theta\right\vert
\geq1$, in the quantum context\ one may find non-singular universe solutions
even in the commutative case. The qualitative\ behavior of the universe
solutions in noncommutative quantum cosmology was discussed in section 4,
where examples were presented that contain non-singular periodic solutions,
non-singular solutions presenting one bounce, as well as non-singular
solutions containing an infinite sequence of bounces enveloped by a larger
bounce. When noncommutativity effects are turned on in the quantum scenario,
they give rise to dynamical universes in situations where Bohmian quantum
cosmology admits only static universes. One example is that of real
wavefunctions, which in the noncommutative FRW model with a conformally
coupled\ scalar field represent always non-singular bouncing universes.
Noncommutativity effects can also induce a dynamical behavior yielding
non-singular periodic universes in cases where the commutative counterpart is
static and the noncommutative wavefunction is complex. An example was
presented in section 6.

The investigation into the NC-C-frame correspondence revealed that, even in
the classical context, the description of the universe evolution provided by
these two possible scenarios for the realization of noncommutativity can
differ radically. In section 3 we showed that for some values of the
integration constants the universe can be non-singular in the NC-frame, while
its C-frame counterpart is singular. An example of the drastic difference that
can occur between the NC- and C-frame realizations was exhibited in that
section for the case where $\theta=\pm1$. It consists of a universe that is
non-singular in the NC-frame and that has no correspondent in the C-frame,
where it is at all times singular.

In the quantum context the distinction between the NC- and C-frame
descriptions was shown to be as relevant as it is in the classical one. An
example was worked out that discusses how a universe with large $a_{nc\min}$
in the NC-frame can correspond to two universes in the C-frame, one with large
$a_{c\min}\simeq$ $a_{nc\min}$, and other with very small $a_{c\min}$. Such an
interplay between small and large scale distances was previously reported in
\cite{22.3}, where it was interpreted as a sort of \textquotedblleft IR-UV
mixing\textquotedblright, in analogy with noncommutative field theory. Other
examples involving the NC-C-frame correspondence were presented and solved
numerically for the case $\left\vert \theta\right\vert <1$ and analytically
for the case $\left\vert \theta\right\vert >1$. Again the case where
$\theta=\pm1$ provided an interesting example. When $\theta=\pm1$
noncommutativity effects\ act to drop the kinetic term from the Wheeler-DeWitt
equation. This justifies the poor and highly constrained dynamics found in
this case. No matter what is the wavefunction in this case, if the initial
conditions are non-singular\ the universe is non-singular and experiments a
single bounce in both NC- and C-frame descriptions (Fig. $5c$).

In case noncommutativity of the minisuperspace variables\ has in fact played a
role in the evolution of the primordial universe, as proposed in \cite{15},
the study carried out in this work renders evident the need of an ontology in
order to have a clear picture of the essential features of the
noncommutative\ universe models. The correspondence between degrees of freedom
in two different frames of realization is not sufficient to define the theory
completely, which is only fixed by assuming one of them as the physical frame.
This necessity seems not to be an exclusive feature of the cosmological model
considered here, where the dramatic difference in the universe evolution can
be attributed, in part, to the fact that the noncommutativity in question is
that of the system's degrees of freedom - the minisuperspace variables-. In
the\ models where noncommutativity does not involve directly the system's
degrees of freedom, as the canonical noncommutative field theories that come
from string theory \cite{3}, the study of the correspondence between the NC-
and C-frame descriptions is also a relevant subject. In the context of gauge
theories, where the connection between the NC- and C- frames is via the
Seiberg-Witten map, an investigation into the properties of the theory that
have resemblance with gravity was carried out, e.g., in \cite{6.5}, where the
equivalence between spacetime\ translations and gauge transformations is shown
to occur in the NC-frame. In the C-frame, on the other hand, where such an
equivalence seems to be lost, noncommutative fields can be interpreted as
ordinary theories immersed in a gravitational background generated by the
gauge field, as shown in the interesting work by Rivelles \cite{30}, and
further in \cite{31}.

\section*{Acknowledgments}

The author acknowledges Nelson Pinto Neto and Jos\'{e} Abdalla
Hela\"{y}el-Neto for corrections in\ earlier versions of this manuscript


\begin{thebibliography}{99}                                                                                               %


\bibitem {1}R. J. Szabo, Phys. Rep. \textbf{378,} 207 (2003).

\bibitem {2}M. R. Douglas and N.A.~Nekrasov, Rev. Mod. Phys. \textbf{73,} 977 (2002).

\bibitem {3}N. Seiberg and E. Witten, J. High Energy Phys. \textbf{09,} 032 (1999).

\bibitem {5}S.~Minwalla, M.~Van~Raamsdonk and N.~Seiberg, J. High Energy Phys.
\textbf{02}, 020 (2000).

\bibitem {6}S. M.~Carroll, J. A.~Harvey, V.A.~Kostelecky, C.D.~Lane and
T.~Okamoto, Phys. Rev. Lett. \textbf{87}, 141601 (2001);

C. E. Carlson, C. D. Carone and R. F. Lebed, Phys. Lett. \textbf{B 518}, 201
(2001); Phys. Lett. \textbf{B 549}, 337 (2002);

A. Anisimov, T. Banks, M. Dine and M. Graesser, Phys. Rev. \textbf{D}
\textbf{65}, 085032 (2002);

J. M. Carmona, J. L. Cort\'{e}s, J. Gamboa and F. M\'{e}ndez, Phys. Lett.
\textbf{B 565}, 222 (2003);

\bibitem {6.5}D. J. Gross and N. A. Nekrasov,\ J.High Energy Phys.
\textbf{10}, 021(2000);

F. Lizzi, R.J. Szabo and A. Zampini, J. High Energy Phys. \textbf{08}, 032 (2001).

\bibitem {11}J. M. Romero and J.A. Santiago, Cosmological Constant and
Noncommutativity: A Newtonian Approach, hep-th/0310266.

\bibitem {12}R. Brandenberger and P.-M. Ho, Phys. Rev. \textbf{D 66}, 023517 (2002);

Q.-G. Huang and M. Li, J. Cosmology Astroparticle Phys. \textbf{11}, 001
(2003); J. High Energy Phys. \textbf{03}, 014 (2003); \textquotedblleft Power
Spectra in Spacetime Noncommutative Inflation\textquotedblright, astro-ph/0311378;

H. Kim, G. S. Lee and Y. S. Myung, ``Noncommutative spacetime effect on the
slow-roll period of inflation'', hep-th/0402018;

H. Kim, G. S. Lee, H. W. Lee and Y. S. Myung,``Second-order corrections to
noncommutative spacetime inflation'', hep-th/0402198;

Y. S. Myung, ``Cosmological parameters in the noncommutative inflation'', hep-th/0407066;

Dao-jun Liu and Xin-zhou Li, ``Cosmological perturbations and noncommutative
tachyon inflation'', astro-ph/0402063;

G. Calcagni,``Noncommutative models in patch cosmology'', hep-th/0406006; ``
Consistency relations and degeneracies in (non)commutative patch inflation'', hep-ph/0406057;

Rong-Gen Cai, Phys. Lett. \textbf{B 593}, 1 (2004);

C.-S. Chu, B. R. Greene and G. Shiu, Mod. Phys. Lett. \textbf{A 16}, 2231 (2001);

S. Tsujikawa, R. Maartens and R. Brandenberger, Phys. Lett. \textbf{B 574},
141 (2003);

H. Kim, G. S. Lee, H. W. Lee and Y. S. Myung, Second-order corrections to
noncommutative spacetime inflation, hep-th/0402198;

G. Calcagni and S. Tsujikawa, Observational constraints on patch inflation in
noncommutative spacetime, astro-ph/0407543.

\bibitem {14}M. Maceda, J. Madore , P. Manousselis e G. Zoupanos, Eur. Phys.
J. \textbf{C 36}, 529 (2004).

\bibitem {15}H. Garcia-Compe\'{a}n, O. Obreg\'{o}n and C. Ram\'{\i}rez, Phys.
Rev. Lett. \textbf{88}, 161301 (2002).

\bibitem {16}G. D. Barbosa and N. Pinto Neto, \textquotedblleft Noncommutative
Geometry and Cosmology\textquotedblright, hep-th/0407111.

\bibitem {17}J. B. Hartle in :High energy Physics 1985, ed. M. J. Bowick and
F. G\"{u}rsey (World Scientific, Singapore, 1986);

J. Feinberg and Y. Peleg, Phys. Rev. \textbf{D 52}, 1988 (1995);

M. Cavagli\`{a}, V. de Alfaro, and A, T. Filippov, Int J. Mod. Phys. \textbf{A
10}, 611 (1995).

\bibitem {18}J. A. de Barros, N. Pinto-Neto, and A. A. Sagioro-Leal, Gen. Rel.
Grav. \textbf{32, }15 (2000).

\bibitem {19}D. Bohm, Phys. Rev. \textbf{85}, 166 (1952); Phys. Rev.
\textbf{85}, 180 (1952).

\bibitem {19.2}D. Bohm, B. J. Hiley and P. N. Kaloyerou, Phys. Rep.
\textbf{144,} 349 (1987).

D. Bohm, B. J. Hiley, The Undivided Universe: An Ontological Interpretation of
Quantum Theory, London (Routledge \& Kegan Paul, 1993);

\bibitem {19.3}S. Goldstein, Phys. Today, March (1998) 42; April (1998) 38.

\bibitem {19.4}J. S. Bell, Speakable and unspeakable in quantum mechanics,
Cambridge University Press, Cambridge, 1993.

\bibitem {20}P. R. Holland, The Quantum Theory of Motion: An Account of the de
Broglie-Bohm Causal Interpretation of Quantum Mechanics (Cambridge University
Press, Cambridge, 1993).

\bibitem {20.3}D. D\"{u}rr. S. Goldstein, and N. Zangh\`{\i}, J. Stat. Phys.
\textbf{67,} 843 (1992).

\bibitem {20.5}D. D\"{u}rr, S. Goldstein, N. Zangh\`{\i}, \textquotedblleft
Bohmian Mechanics and the Meaning of the Wave Function\textquotedblright,
\textquotedblleft Experimental Metaphysics-Quantum Mechanical Studies in Honor
of Abner Shimony,\textquotedblright\ ed. R.S.Cohen, M. Horne, and J. Stachel,
Boston Studies in the Philosophy of Science (Kluwer, 1996) [quant-ph/9512031].

\bibitem {20.6}D. D\"{u}rr , S. Goldstein and N. Zangh\`{\i},
\textquotedblleft Bohmian Mechanics as the Foundation of Quantum
Mechanics\textquotedblright, Contribution to \textquotedblleft Bohmian
Mechanics and Quantum Theory: An Appraisal,\textquotedblright\ edited by J.T.
Cushing, A. Fine, S. Goldstein, Kluwer Academic Press.

\bibitem {20.7}V. Allori and N. Zangh\`{\i}, \textquotedblleft What is Bohmian
Mechanics\textquotedblright, quant-ph/0112008.

\bibitem {21}H. Nikolic, \textquotedblleft Covariant canonical quantization of
fields and Bohmian mechanics\textquotedblright, hep-th/0407228;

D. Durr, S. Goldstein, R. Tumulka, N. Zanghi, J. Phys. \textbf{A 36, }4143 (2003);

C. Colijn, E. R. Vrscay, Phys. Lett. \textbf{A 300,} 334 (2002);

A. S. Sans, F. Borondo and S. Miret-Art\'{e}s, J. Phys. Condens. Matter
\textbf{14,} 6109 (2002);

D. Home and A. S. Majumdar, Found. Phys. \textbf{29}, 721 (1999);

L. Delle Site, Europhysics Letters \textbf{57}, 20 (2002);.

G. Gruebl, R. Moser, K. Rheinberger,\ J. Phys. \textbf{A 34}, 2753 (2001).

Md. Manirul Ali, A. S. Majumdar and D. Homel, Phys. Lett. \textbf{A 334,}
61\textbf{ }(2002);

E. Guay and L. Marchildon, J. Phys. \textbf{A 36} 5617 (2003);

D. D\"{u}rr, S. Goldstein, S. Teufel and N. Zanghi, Physica \textbf{A 279},
416 (2000);

\bibitem {22}J. Kowalski-Glikman and J. C. Vink,\ Class. Quant.
Grav.\ \textbf{7,} 901 (1990);

Squires, Phys. Lett. \textbf{A 162,} 35 (1992);

Y. V. Shtanov, Phys. Rev. \textbf{D 54,} 2564 (1996);

R. Colistete Jr., J. C. Fabris, N. Pinto-Neto, Phys. Rev. \textbf{D 62,} 083507 (2000);

F. Shojai, A. Shojai, J. High Energy Phys. \textbf{05, }037 (2001); Class.
Quant. Grav.\ \textbf{21}, 1 (2004);

A. S. Sans, F. Borondo and S. Miret-Art\'{e}s, J. Phys. Condens. Matter
\textbf{14}, 6109 (2002);

W.-H. Huang, I.-C. Wang, \textquotedblleft Quantum Perfect-Fluid Kaluza-Klein
Cosmology\textquotedblright, gr-qc/0309042;

J. Acacio de Barros and N. Pinto-Neto,\ Phys. Lett. \textbf{A 241}, 229 (1998);

R. Colistete Jr., J. C. Fabris, N. Pinto-Neto, Phys. Rev. \textbf{D 62,} 083507 (2000);

N. Pinto-Neto, E. S. Santini, Phys. Rev. \textbf{D 59,} 123517 (1999).

\bibitem {22.1}C. Kiefer, \textquotedblleft Conceptual issues in quantum
cosmology\textquotedblright, Lect. Notes Phys. 541 (2000) [gr-qc/9906100];

N. Pinto Neto, \textquotedblleft Quantum Cosmology\textquotedblright, VIII
Brazilian School of Cosmology and Gravitation (Editions Frontieres,
Gif-sur-Yvette, 1996).

\bibitem {22.5}V. Faraoni, E. Gunzig and P. Nardone, Fund. Cosmic Phys.
\textbf{20,} 121\ (1999).

\bibitem {22.2}G. D. Barbosa, J. High Energy Phys. \textbf{05}, 024 (2003).

\bibitem {22.3}G. D. Barbosa and N. Pinto-Neto, Phys. Rev. \textbf{D 69},
065014 (2004).

\bibitem {22.7}M.~Chaichian, M. M. Sheikh-Jabbari and A. Tureanu, Phys. Rev.
Lett. \textbf{86}, 2716 (2001).

\bibitem {27}L. M. Campbell and L. J. Garay, Phys, Lett. \textbf{B 254}, 49 (1991);

M. Cavagli\`{a}, Mod. Phys. Lett.\textbf{ A 9}, 1897 (1994).

\bibitem {28}C. Simeone, Gen. Rel. Grav. \textbf{34}, 1887 (2002).

\bibitem {29}A. Peres and D. R. Terno, Rev. Mod. Phys. \textbf{76}, 93\ (2004).

\bibitem {26.5}I. S. Gradshteyn and I. M. Ryzhik, Table of Integrals, Series
and Products, Academic Press, San Diego, fourth edition.

\bibitem {30}V. O. Rivelles, Phys. Lett. \textbf{B 558,} 191 (2003).

\bibitem {31}Hyun Seok Yang, \textquotedblleft Exact Seiberg-Witten Map and
Induced Gravity from Noncommutativity\textquotedblright, hep-th/0402002.
\end{thebibliography}
\end{document}